\documentclass[11pt,a4paper,fleqn]{article}
\usepackage{jcappub}
\usepackage{graphicx}
\usepackage{amsmath}
\usepackage{mathtools}
\usepackage{multirow}
\usepackage{hyperref}
\usepackage{soul}
\usepackage{color}
\usepackage{rotating}
\usepackage[english]{babel}
\usepackage{float}
\usepackage{cleveref}
\usepackage{hhline}
\usepackage{enumerate}
\usepackage{makecell}
\usepackage{subfigure}

\title{Measuring the cosmic dipole with golden dark sirens in the era of next-generation ground-based gravitational wave detectors}
\date{\today}

\author[a]{Anson Chen}
\affiliation[a]{International Center for Theoretical Physics (Asia-Pacific), University of Chinese Academy of Sciences, No.80 Zhongguancun East St., Beijing, China}
\emailAdd{chena@ucas.ac.cn}

\abstract{The tensions between cosmological parameter measurements from the early-universe and the late-universe datasets offer an exciting opportunity to explore new physics, if not accounted for unknown systematics. Apart from the well-known Hubble tension, a tension up to $\sim4.9\sigma$ in the cosmic dipole has also been reported. While the cosmic dipole is mainly induced by the observer’s kinetic motion, an intrinsic dipole arising from the anisotropy of the universe could also play an import role. Such an intrinsic
anisotropy can be a dark energy mimicker that causes the observed accelerating expansion of the universe. As a new and powerful tool, gravitational waves can serve as an independent probe to the cosmic dipole. A useful type of events to achieve this is the “golden dark sirens”, which are near-by well-localized compact binary coalescences whose host galaxies can be identified directly due to precise localization. By forecasting golden dark sirens obtained from 10-year observations using different possible detector networks in the future, we find that the standard LIGO-Virgo-KAGRA detectors are not able to detect a meaningful amount of golden dark sirens, and hence next-generation ground-based detectors are essential to obtain a strong constraint on the cosmic dipole. In particular, we find that a three-detector network consisting of more than one next-generation detectors can yield a constraint on the cosmic dipole at an order of $10^{-3}$ when jointly measured with $H_0$. Moreover, a constraint on the cosmic dipole at an order of $10^{-4}$ can be achieved when fixing $H_0$.}

\arxivnumber{2505.12678}

\begin{document}
\maketitle

\section{Introduction}

The Lambda Cold Dark Matter ($\Lambda$CDM) model is an excellent and concise fit to the cosmic microwave background (CMB) dataset from the Planck mission \cite{Planck:2018vyg}, which offers strict constraints on cosmology parameters. However, datasets from late universe observations give rise to several tensions to the measurements of some cosmological parameters, such as the Hubble tension and the $S_8$ tension (see review in \cite{Abdalla:2022yfr}). The Hubble tension between measurement from the Planck mission ($H_0=67.49\pm0.50~{\rm km~s}^{-1}{\rm Mpc}^{-1}$) \cite{Planck:2018vyg} and that by using Cepheids and Type Ia supernovae (SNIa) as standard candles by the SH0ES collaboration ($H_0=74.03\pm1.42~{\rm km~s}^{-1}{\rm Mpc}^{-1}$) \cite{Riess:2019cxk} has reached $\sim5\sigma$. If such a discrepancy is not caused by unknown systematic errors, it is most likely a hint to new physics beyond the $\Lambda$CDM model.

In addition, a tension in the cosmic dipole also emerges from observations in the local universe compared to the CMB measurement. The cosmic dipole consists of two parts, namely the kinetic dipole and the intrinsic dipole. The kinetic dipole is induced by the motion of the solar system with respect to the CMB rest frame. Its most precise measurement comes from the Planck satellite, which yields an amplitude of $(1.23\pm0.00036)\times10^{-3}$ and a direction of $(264.02^\circ\pm0.01^\circ,48.253^\circ\pm0.005^\circ)$ in galactic coordinates \cite{Planck:2018nkj}. On the other hand, the intrinsic dipole could arise from the violation of the cosmology principle of isotropy and homogeneity. Such an anisotropy may be resulted from backreaction of cosmological perturbations, and can resemble the effects of dark energy \cite{Kolb:2005da,Bolejko:2016qku}. Although its origin is not yet fully understood, the observed cosmic dipole tension might reveal a hint of the existence of an intrinsic dipole. For example, the number count dipole inferred from quasar counting in the CatWISE2020 catalogue yields an amplitude of $1.5\times10^{-2}$, which results in a tension of $4.9\sigma$ to the CMB solar dipole \cite{Secrest:2020has}. In addition, the number count dipole amplitude measured from radio galaxy catalogues, such as the TIFR GMRT Sky Survey (TGSS), the NRAO VLA Sky Survey (NVSS), and the Westerbork Northern Sky Survey (WENSS), ranges from $0.010$ to $0.070$ with an uncertainty in the order of $10^{-3}$ \cite{Blake:2002gx,2011ApJ...742L..23S,Gibelyou:2012ri,Tiwari:2013vff,Fernandez-Cobos:2013fda,Tiwari:2013ima,Bengaly:2017slg,Siewert:2020krp}, which are several times larger than the CMB results. Moreover, there also exists a $3.3\sigma$ tension in the dipole amplitude measured with SNIa surveys \cite{Singal:2021crs}. Although some efforts have been made to alleviate the cosmic dipole tension, such as associating the local dipole with the impact from local structure \cite{Tiwari:2015tba,Colin:2017juj}, or explaining it by the existence of a large void in the local universe \cite{Rubart:2014lia}, they cannot fully reconcile the tension. In addition, a time evolution model of radio source population is proposed to ease the tension \cite{Dalang:2021ruy}, but it has been disapproved by recent works \cite{vonHausegger:2024jan,vonHausegger:2024fcu}. Apart from these, attempts to measure the intrinsic dipole have also been made \cite{Ferreira:2020aqa,Ferreira:2021omv}. Nevertheless, a dipole consistent to the CMB value has recently been obtained with a novel tomographic approach using eBOSS data in \cite{daSilveiraFerreira:2024ddn}, which is in tension with other number count studies.

On the other hand, the use of gravitational waves (GWs) from compact binary coalescences (CBCs) in studying cosmology has been rapidly developed in the past few years. The first binary neutron star (BNS) merger GW170817 accompanied by electromagnetic (EM) counterparts gives an estimation of the Hubble constant $H_0=70^{+12}_{-8}~{\rm km~s}^{-1}{\rm Mpc}^{-1}$ \cite{LIGOScientific:2017adf}. Such a GW event is called a bright siren, whose host galaxy can be identified with follow-up observations of EM signals. However, a second bright siren hasn't been detected in observation runs of the LIGO-Virgo-KAGRA (LVK) detector network since then. Alternatively, a Bayesian framework to measure $H_0$ with a set of dark sirens, which are GW events without EM counterparts such as binary black hole (BBH) and neutron-star-black-hole (NSBH) mergers, has become a mature technique \cite{Schutz1986,Finke:2021aom,Mastrogiovanni:2021wsd,Mastrogiovanni:2023emh,Mastrogiovanni:2023zbw,Gray:2019ksv,Gray:2021sew,Gray:2023wgj}. In such analysis, the redshift prior of GW events is constructed with compact object population and merger rate information, as well as galaxy redshifts within GW localization areas obtained from galaxy catalogues. By combining the bright siren result with the dark siren analysis using 42 selected BBHs from the GWTC-3 catalogue, a better constraint of $H_0=68^{+8}_{-6} {\rm km~s^{-1}Mpc^{-1}}$ is obtained \cite{LIGOScientific:2021aug}.

Although the number of events detected by LVK has reached an order of $10^2$ today, the current data is still insufficient to probe the cosmic dipole with number counting or BBH mass distribution \cite{Stiskalek:2020wbj,Essick:2022slj,Kashyap:2022ibx}. However, some forecasts have been made to measure the cosmic dipole using the next-generation ground-based or space-based detectors, such as the Einstein Telescope (ET) \cite{Punturo:2010zz,Branchesi:2023mws}, the Cosmic Explorer (CE) \cite{Evans:2021gyd,Srivastava:2022slt,Evans:2023euw}, and the Laser Interferometer Space Antenna (LISA) \cite{amaroseoane2017,Bayle:2022hvs}. Since ET and CE are expected to detect CBC signals in an order of $10^5$-$10^6$ per year, measurements of the cosmic dipole with CBC number counting in the era of next-generation detectors have been discussed in ref. \cite{Mastrogiovanni:2022nya,Grimm:2023tfl}. In addition, measurements with mock bright sirens detected by different future detector networks are performed in \cite{Cousins:2024bhk}, where an order of $10^3$ bright sirens are forecasted with the network of 1ET+2CE per year. Furthermore, the potential in dipole measurements for bright sirens in decihertz or millihertz frequency bands from massive BBH mergers has been analyzed in \cite{Cai:2017aea,Cai:2019cfw}. On the other hand, measurements of the kinetic dipole from stochastic GW background by pulsar timing arrays have also been proposed in \cite{Cusin:2022cbb,Tasinato:2023zcg,Cruz:2024svc,Cruz:2024diu}.

Despite the accurate measurement of the cosmic dipole with mock bright sirens in \cite{Cousins:2024bhk,Cai:2017aea,Cai:2019cfw}, there is a concern that the original luminosity distances ($D_L^0$) of GW events in the CMB rest frame are assumed to be known in these analyses, and the dipole measurement is obtained by $\chi^2$ minimization between observed $D'_L$ and theoretical $D_L$ constructed by $D_L^0$ with dipole modification. While in reality, $D_L^0$ cannot be obtained by observations. Even though EM counterparts of bright sirens can be used to pin down their host galaxies, the galaxy redshifts are also modified by the cosmic dipole, and hence cannot be converted to $D_L^0$ directly. Moreover, the detection rate of bright sirens are still not fully understood, which may undermine the measurability of the cosmic dipole with bright sirens. Therefore in this work, we propose a more robust measurement of the cosmic dipole using ``golden" dark sirens, which are highly well-localized CBCs at redshift $z<0.1$ so that only one brightest galaxy can be found within the CBC localization area.  This brightest galaxy is then considered to be the host galaxy of the CBC, from which its redshift can be determined with follow-up EM observations \cite{Singer:2016eax}. Highly precise measurements of $H_0$ with golden dark sirens have been forecasted in various works \cite{Nishizawa:2016ood,Yu:2020vyy,Borhanian:2020vyr,Gupta:2022fwd,Chen:2024gdn}. Because of the promising prospect to obtain their redshifts, we will thus forecast the cosmic dipole measurability of golden dark sirens detected by different future detector networks consisting of ET, CE and LVK with A+ or A\# sensitivity~\cite{A_sharp_report}.

On the other hand, since the cosmic dipole changes the observed values of $D_L$ for CBCs, it may cause a bias in $H_0$ measurement. Furthermore, for the dark siren method, the improvement in $H_0$ constraint relies on increasing the event number. However, the observed event number density would also vary in different regions on the sky due to the cosmic dipole effects. As a result, a bias in $H_0$ constraint may also be induced by uneven event numbers in different parts of the sky when combining posteriors for all events. Therefore we will also investigate such an effect from the cosmic dipole in the dark siren method in this work.

The structure of this paper is presented as followed. In Sec. \ref{sec:cosmo_effects}, we first investigate the effects of the cosmic dipole in $H_0$ measurement with the dark siren method. Then in Sec. \ref{sec:measurements}, we introduce the approach to jointly constrain $H_0$ and the cosmic dipole using golden dark sirens forecasted with different detector networks. The results are then shown in Sec. \ref{sec:results}. Finally, Sec. \ref{sec:conclusions} summarizes the findings of this work.

\section{Cosmic dipole effects in dark siren cosmology}
\label{sec:cosmo_effects}

In this section we will analyze the effects due to the presence of the cosmic dipole in $H_0$ measurement with the dark siren method. We use mock GW event simulation to forecast $H_0$ constraint from data in the fourth and the fifth observation run (O4 and O5) of the LVK, and examine whether there exists biases induced by the cosmic dipole.

\subsection{The cosmic dipole}

We denote the cosmic dipole amplitude along the line-of-sight $\hat{z}(\phi,\theta)$ as $g(\hat{n}\cdot\hat{z})$, where $g$ denotes the magnitude of the dipole, and $\hat{n}(\phi^{\rm dip},\theta^{\rm dip})$ is the direction of the dipole given by
\begin{equation}
    \hat{n} = (\cos\phi^{\rm dip}\sin\theta^{\rm dip}, \sin\phi^{\rm dip}\sin\theta^{\rm dip}, \cos\theta^{\rm dip}).
\end{equation}
Here $\phi^{\rm dip}$ and $\theta^{\rm dip}$ are the right ascension and the declination of the dipole direction. For the kinetic part of the cosmic dipole, $g^{\rm kin}=-|\vec{v}_o|/c$, where $\vec{v}_o$ is the observer velocity with respect to the CMB rest frame. Note that the dipole direction is opposite to the observer velocity direction by our definition.

The observed luminosity distance of a source is modified by the cosmic dipole effects as \cite{Bonvin:2005ps}
\begin{equation}
    D_L^{\rm obs} = D_L^0[1+g(\hat{n}\cdot\hat{z})].
    \label{eq:DL_dip}
\end{equation}
Meanwhile the observed redshift of the source is also modified by
\begin{equation}
    1+z^{\rm obs} = (1+z^0)[1+g(\hat{n}\cdot\hat{z})],
    \label{eq:z_dip}
\end{equation}
where $D_L^0$ and $z^0$ are the luminosity distance and the redshift in the CMB rest frame. In the local universe we have 
\begin{equation}
	D_L^0(z^0) = \frac{c(1+z^0)}{H_0}\int_0^{z^0} \frac{{\rm d}z'}{\sqrt{\Omega_{m0}(1+z')^3+\Omega_{\Lambda 0}}},
\end{equation}
where $\Omega_{m0}$ and $\Omega_{\Lambda 0}$ are energy density of matter and dark energy today.

Moreover, since the observed mass of a CBC from a GW signal is the redshifted mass, its relation to the source-frame mass $m^{\rm s}$ also becomes
\begin{equation}
    m^{\rm obs} = m^{\rm s}(1+z^{\rm obs}) = m^{\rm s}(1+z^0)[1+g(\hat{n}\cdot\hat{z})].
    \label{eq:m_dip}
\end{equation}
Since the dark siren method uses information from the black hole mass distribution model, the cosmic dipole effects in observed component masses of CBCs may lead to systematics in $H_0$ constraints.

\subsection{Bayesian framework for dark siren cosmology}

In the hierarchical Bayesian formalism \cite{Fishbach:2018edt,Mandel:2018mve,Vitale:2020aaz}, the posterior probability of a set of hyper-parameters $\Lambda$ given a set of GW data $\{x\}$ from $N_{\rm obs}$ observed events reads 
\begin{equation}
    p(\Lambda|\{x\}) \propto {\cal L}(\{x\}|N_{\rm obs},\Lambda)p(N_{\rm obs}|\Lambda) \pi(\Lambda),
\end{equation}
where $\Lambda$ includes parameters in the population model, merger rate evolution, and the cosmology model. The term $\pi(\Lambda)$ is the prior on $\Lambda$, and $p(N_{\rm obs}|\Lambda)$ is the probability of observing $N_{\rm obs}$ events out of an expected number of observations $N_{\rm exp}$, which can be described by a Poission distribution. By choosing a prior $\pi(N_{\rm exp})\propto 1/N_{\rm exp}$, it can be marginalized over in the following analysis \cite{Fishbach:2018edt}.

The term ${\cal L}(\{x\}|N_{\rm obs},\Lambda)$ is the likelihood of GW data $\{x\}$ form $N_{\rm obs}$ GW events, which can be expanded as \cite{Mastrogiovanni:2021wsd}
\begin{equation}
	{\cal L}(\{x\}|N_{\rm obs},\Lambda) = \prod^{N_{\rm obs}}_i \frac{\int p(x_i|\theta_p,\Lambda)p_{\rm pop}(\theta_p|\Lambda){\rm d}\theta_p}{\int p_{\det}(\theta_p,\Lambda)p_{\rm pop}(\theta_p|\Lambda){\rm d}\theta_p}.
	\label{eq:bayesian_likelihood}
\end{equation}
Here $p(x_i|\theta_p,\Lambda)$ is the likelihood of GW data form a detected event given GW signal parameters $\theta_p$ and hyper-parameters $\Lambda$, and $p_{\rm pop}(\theta_p|\Lambda)$ is the prior on $\theta_p$ given a population model. The term $p_{\det}(\theta_p,\Lambda)$ denotes the detection probability of GW data given selection criteria such as the signal-to-noise (SNR, $\rho$) threshold ($\rho_{\rm th}$), which is computed by
\begin{equation}
	p_{\det}(\theta_p,\Lambda) = \int_{\rho>\rho_{\rm th}} p(x_i|\theta_p,\Lambda)p_{\rm pop}(\theta_p|\Lambda) {\rm d}x_i .
\end{equation}

At the moment, there are several Bayesian framework code packages for dark siren cosmology, for example \texttt{ICAROGW} \cite{Mastrogiovanni:2021wsd,Mastrogiovanni:2023emh,Mastrogiovanni:2023zbw}, \texttt{GWCOSMO} \cite{Gray:2019ksv,Gray:2021sew,Gray:2023wgj} and \texttt{CHIMERA} \cite{Borghi:2023opd,Tagliazucchi:2025ofb}. They both include population analysis and galaxy catalogue supports for cosmological measurements. However, the current GLADE+ galaxy catalogue \cite{Dalya:2021ewn} used in LVK cosmology analysis is highly incomplete at large distances ($D_L>130$ Mpc). Therefore, current galaxy catalogue supports in $H_0$ measurements are very limited for LVK events (see e.g. \cite{Gray:2023wgj,Chen:2023wpj}). Hence, to avoid increased computational costs form associating with galaxt catalogues, we use \texttt{ICAROGW} v1.0 \cite{Mastrogiovanni:2021wsd} in this work to demonstrate the effects of the cosmic dipole in $H_0$ measurements, which only uses the population method, such that
\begin{equation}
	p_{\rm pop}(\theta_p|\Lambda) \propto p(m_1^s,m_2^s|\Lambda)\frac{{\rm d}V_c}{{\rm d}z}(\Lambda)p_{\rm rate}(z|\Lambda).
\end{equation}
Here $p(m_1^s,m_2^s|\Lambda)$ is the phenomenological compact object mass distribution in the source frame. The term ${\rm d}V_c/{\rm d}z$ implies that the number of CBCs grows with the comoving volume $V_c$ in terms of redshift. The last term $p_{\rm rate}(z|\Lambda)$ describes the merger rate evolution in redshift, which is assumed to be analogous to the Madau star formation rate redshift evolution model \cite{Madau:2014bja,Callister:2020arv} as
\begin{equation}
	p_{\rm rate}(z|\Lambda) = R_0 [1+(1+z_p)^{-\gamma-k}] \frac{(1+z)^\gamma}{1+\left(\frac{1+z}{1+z_p}\right)^{\gamma+k}}.
\end{equation}
From GWTC-3 analysis, a powerlaw mass distribution with the addition of a Gaussian peak is favoured for the black hole population model $p(m_1^s,m_2^s|\Lambda)$ \cite{KAGRA:2021duu}, which contains 8 parameters.

\subsection{GW event simulation}
\label{sec:GW_simulation}

We use the \texttt{GWSim} package \cite{Karathanasis:2022hrb} to simulate detected GW events with different detector networks and sensitivities. \texttt{GWSim} encodes the compact object population models and merger rate redshift evolution used in LVK data analysis. Its original version contains detector sensitivities up to the LVK O4 stage\footnote{L1 and H1: https://git.ligo.org/publications/detectors/obs-scenarios-2019/blob/master/\\CurvesForSimulation/aligo\_O4high.txt\\
V1: https://git.ligo.org/publications/detectors/obs-scenarios-2019/blob/master/CurvesForSimulation/\\avirgo\_O4high\_NEW.txt\\
K1: https://git.ligo.org/publications/detectors/obs-scenarios-2019/blob/master/CurvesForSimulation/\\kagra\_80Mpc.txt}. We modify it to include the LVK O5 (A+)\footnote{L1 and H1: https://dcc.ligo.org/DocDB/0149/T1800042/005/AplusDesign.txt\\
V1: https://git.ligo.org/publications/detectors/obs-scenarios-2019/blob/master/Scripts/Figure1/data/\\avirgo\_O5high\_NEW.txt\\
K1: https://git.ligo.org/publications/detectors/obs-scenarios-2019/blob/master/CurvesForSimulation/\\kagra\_128Mpc.txt} and A\#\footnote{https://dcc.ligo.org/LIGO-T2300041-v1/public/Asharp\_strain.txt} sensitivities. We also include the ET noise curve with 10 km arm length\footnote{https://apps.et-gw.eu/tds/ql/?c=16492/ET10kmcolumns.txt} and the CE noise curve with 40 km arm length\footnote{https://dcc.cosmicexplorer.org/CE-T2000017/public/cosmic\_explorer\_strain.txt}. We assume that ET is an L-shape detector with the same location and orientation as Virgo, and CE has the same location and orientation as LIGO Livingston or Hanford based on the number of CE detectors. 

With input cosmological parameters $H_0$ and $\Omega_{m0}$, \texttt{GWSim} generates mock galaxy distribution with a constant number density, and creates CBC events based on input population and merger rate evolution parameters whose host galaxies are randomly assigned to the mock galaxies for each redshift bin. Then the SNR of each event is computed by
\begin{equation}
	\rho = \left[4\int^{\infty}_0 \frac{|\tilde{h}(f)|^2}{S_n(f)} \right]^{1/2},
\end{equation}
where $\tilde{h}(f)$ is the GW waveform in the frequency domain, and $S_n(f)$ is the spectral noise density of the detector. The detected events are selected with SNRs higher than the chosen threshold. The duty cycle for each detector is set to be 0.75.

To forecast the effects of the cosmic dipole in dark siren analysis in the LVK O4 stage, we generate mock O4 events for an observation time of 3 years. We simulate a mock universe with injected $H_0=70~{\rm km~s}^{-1}{\rm Mpc}^{-1}$ and $\Omega_{m0}=0.3$, and the galaxy number density of $10^{-5}~{\rm Mpc}^{-3}$. We use the Powerlaw + Peak black hole population model and the Madau merger rate redshift evolution model as used in the GWTC-3 cosmology paper \cite{LIGOScientific:2021aug}, with the parameters listed in Table \ref{tab:param}. We first generate mock events without injecting a cosmic dipole. The number of detections is 206 for an SNR threshold of 11. We also generate mock O5 events for 1 year of observations with the same setup, and obtain 3062 detections. Because the computational costs are high for the current \texttt{ICAROGW} package to handle thousands of events, we only use it to analyze 1 year of O5 detections, and we will approximate the analysis for 5 year of O5 detections in Sec. \ref{subsec:H0_effects}.

\begin{table}[t]
    \centering
    \begin{tabular}{c|c|c}
        \hline
        Parameter & Definition & Value \\
        \hline
       $\alpha$  & \makecell{The power of the power-law\\primary mass distribution.}  & $3.78$ \\
       \hline
       $\beta$  & \makecell{The power of the power-law\\mass ratio distribution.}  & $0.81$ \\
       \hline
       $m_{\rm min}^{\rm BH}$ & \makecell{The minimum mass of the\\mass distribution $[M_\odot]$.} & $4.98$ \\
       \hline
       $m_{\rm max}^{\rm BH}$ & \makecell{The maximum mass of the\\mass distribution $[M_\odot]$.} & $112.5$ \\
       \hline
       $\lambda_p$ & \makecell{Fraction of the mass model in\\the Gaussian component.} & $0.03$ \\
       \hline
       $\mu_g$ & \makecell{Mean of the Gaussian component in\\the primary mass distribution.} & $32.27$ \\
       \hline
       $\sigma_g$ & \makecell{Width of the Gaussian component in\\the primary mass distribution.} & $3.88$ \\
       \hline
       $\delta_m$ & \makecell{Range of mass tapering at the lower\\end of the mass distribution.} & $4.8$ \\
       \hline
       $R_0$  & \makecell{The merger rate at $z=0$\\$[{\rm yr}^{-1}{\rm Gpc}^{-3}]$.}  & $20$ \\
       \hline
       $\gamma$  & \makecell{The power of the power-law rate\\evolution before redshift $z_p$.}  & $4.59$ \\
       \hline
       $k$  & \makecell{The power of the power-law rate\\evolution after redshift $z_p$.}  & $2.86$ \\
       \hline
       $z_p$  & \makecell{The redshift turning point of\\the rate evolution.} & $2.47$ \\
       \hline
    \end{tabular}
    \caption{Parameters in the Power Law + Peak black hole mass distribution model and the Madau merger rate redshift evolution used in our GW simulation. The values are adopted from the LVK GWTC-3 cosmology paper \cite{LIGOScientific:2021aug}.}
    \label{tab:param}
\end{table}

\begin{figure*}[t]
    \centering
    \includegraphics[width=\textwidth]{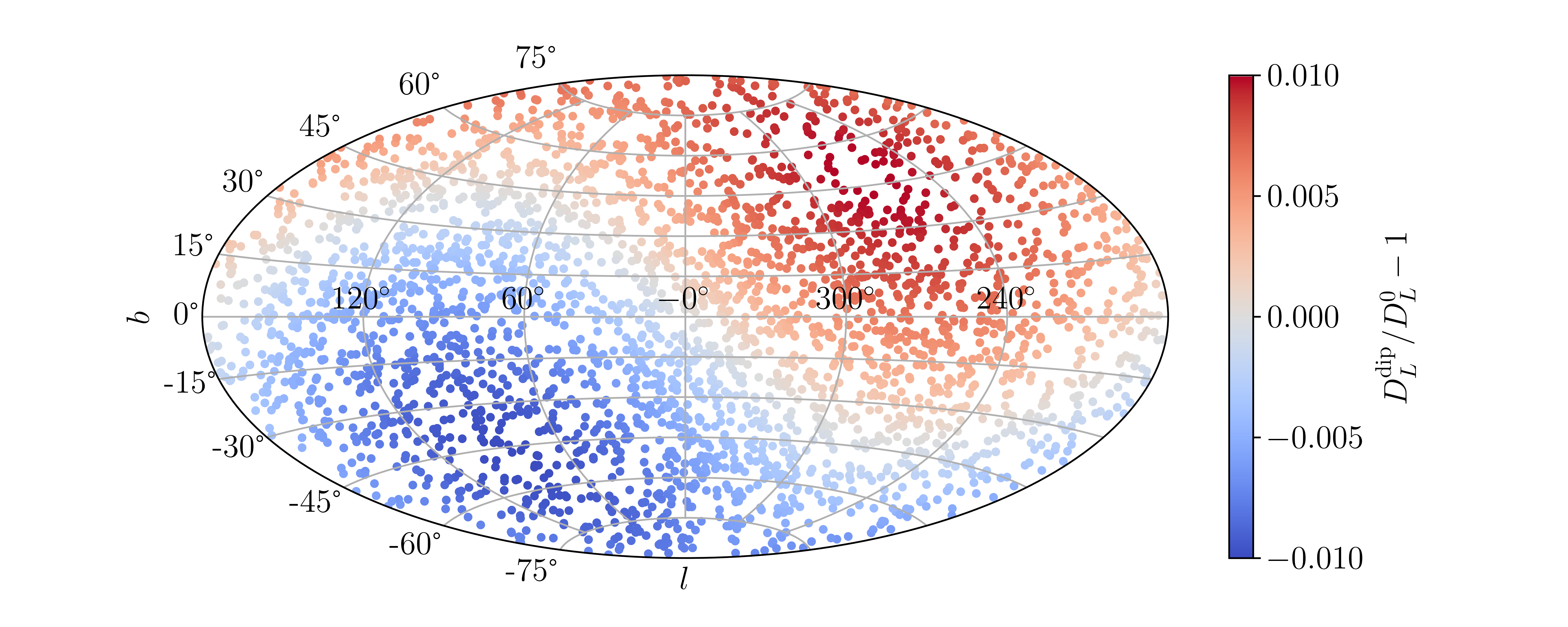}
    \caption{Skymap showing the detected GW events for 1 year of LVK O5 observations in galactic coordinates. The color indicates the modification in $D_L^{\rm obs}$ due to a cosmic dipole with $g=0.01$ and $(l^{\rm dip},b^{\rm dip})=(264^\circ,48^\circ)$.}
    \label{fig:skymap_O5}
\end{figure*}

Next we use the same random seed to generate mock GW signals in the presence of a cosmic dipole. Now the signals are generated with modified observed component masses $m_1^{\rm obs}$, $m_2^{\rm obs}$ and luminosity distance $D_L^{\rm obs}$ computed by eq.~\eqref{eq:m_dip} and \eqref{eq:DL_dip}. The detected events are then selected by the SNRs of these new signals. Fig. \ref{fig:skymap_O5} shows the modification to $D_L^{\rm obs}$ for detected events from 1 year of O5 observations due to the cosmic dipole with the direction measured from the CMB. We find that the detection number of events increases by 11 with a dipole magnitude $g=0.01$.

\subsection{Mock posterior samples}

Once we obtain mock GW events, we generate posterior samples for parameters of the GW signals using \texttt{GWDALI} \cite{deSouza:2023ozp}, a package to perform derivative approximation for likelihood. This approximation is based on Taylor expansion of the logarithm of the likelihood around the maximum likelihood
value ${\cal L}_0$, where the first order term of derivatives corresponds to the Fisher matrix approximation:
\begin{align}
	\log {\cal L} \approx & \log {\cal L}_0 - \frac{1}{2}\sum_{i,j}\langle\partial_i h(\theta_p)|\partial_j h(\theta_p)\rangle_0 \Delta\theta_{p,i} \Delta\theta_{p,j} \nonumber\\
	& + {\cal O}(\partial^2).
\label{eq:DALI}
\end{align}
Here the scalar product is defined as
\begin{equation}
	\langle h_i|h_j \rangle = 2\int^{\infty}_0 \frac{\tilde{h}_i(f)\tilde{h}^*_j(f)+\tilde{h}^*_i(f)\tilde{h}_j(f)}{S_n(f)} {\rm d}f,
\end{equation}
\texttt{GWDALI} generates stochastic samples for $\theta_p$ by the approximated likelihood computed in eq.~\eqref{eq:DALI}. In this way we can generate posterior samples for many mock events at a much lower computational cost than using the full likelihood.

We use the IMRPhenomXPHM waveform \cite{Pratten:2020ceb} to generate posterior samples for 6 GW parameters $\{m_1^{\rm obs},m_2^{\rm obs},D_L^{\rm obs},\iota,\phi,\theta\}$, where $\iota$ is the inclination angle of the CBC, $\phi$ and $\theta$ are the right ascension and the declination of the signal. By including the higher modes in the IMRPhenomXPHM waveform, the degeneracy between $D_L^{\rm obs}$ and $\iota$ can be reduced. The posterior samples are generated for mock events each time we change the injected cosmic dipole magnitude.

\subsection{Effects in $H_0$ constraint}
\label{subsec:H0_effects}

With posterior samples for GW parameters at hand, we then compute the posterior probability of $H_0$ using \texttt{ICAROGW}. We generate $10^6$ GW injections to compute the detection probability in the denominator of eq.~\eqref{eq:bayesian_likelihood}. We fix the population model and the merger rate redshift evolution parameters, and compute the 1-dimension posterior distribution of $H_0$. After the posterior distribution for each event is computed, we multiply them to obtain a combined posterior distribution $p(H_0)$, because there is no correlation between individual events for 1-dimension posterior computation. Such straightforward multiplication is not applicable in multi-dimensional parameter space where parameters correlate with each other.

\begin{figure}[t]
    \centering
    \includegraphics[width=0.6\linewidth]{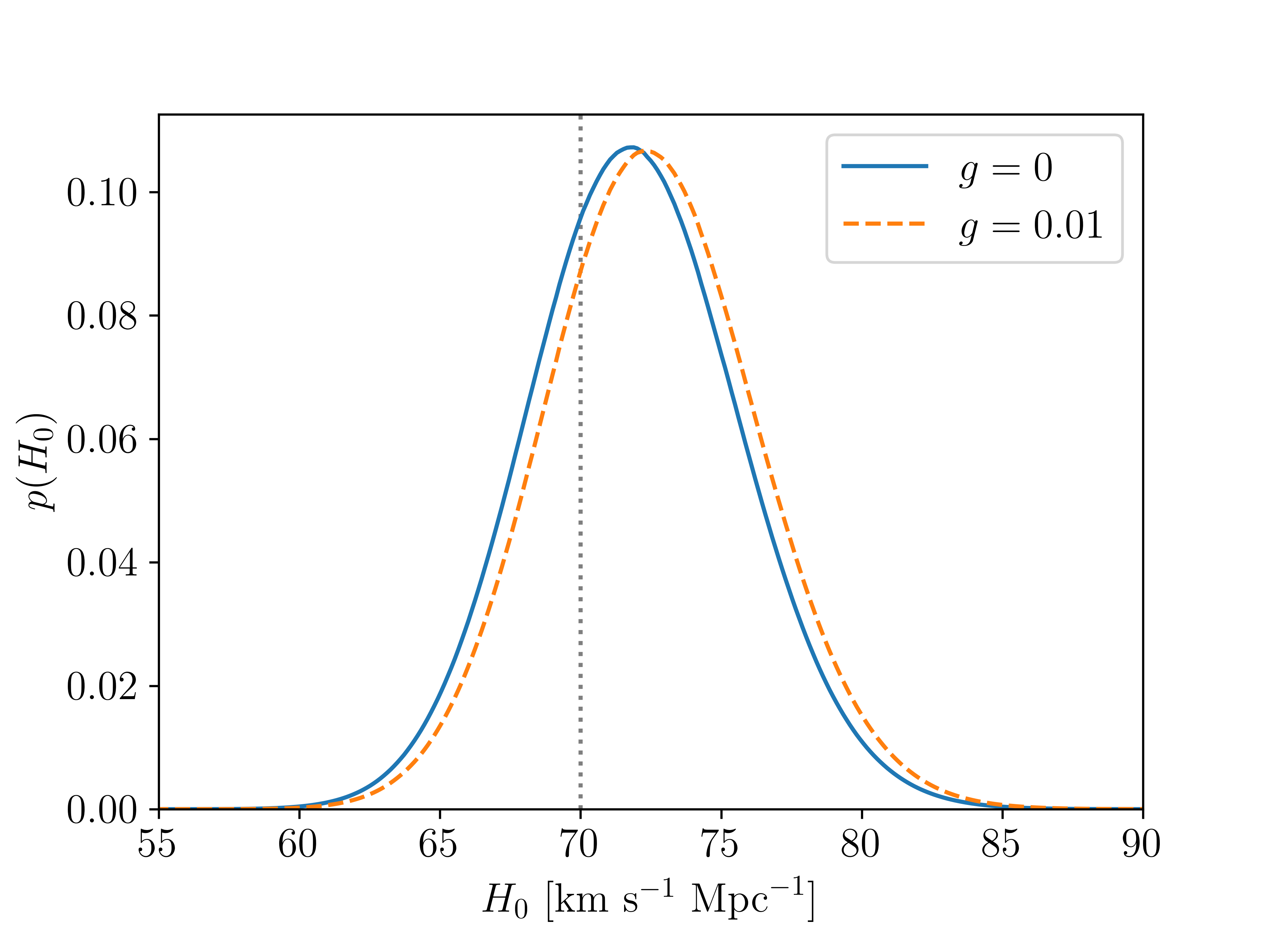}
    \caption{The comparison between combined posterior distribution for $H_0$ using 206 mock O4 events with and without injecting a cosmic dipole. The grey line shows the injected value of $H_0=70~{\rm km}~{\rm s}^{-1}~{\rm Mpc}^{-1}$.}
    \label{fig:pH0_dip_O4}
\end{figure}

We first compute $p(H_0)$ using 206 mock O4 events without injecting a cosmic dipole, and obtain $H_0=71.85^{+3.65}_{-3.81}~{\rm km}~{\rm s}^{-1}~{\rm Mpc}^{-1}$ with $1\sigma$ uncertainty, as shown in Fig. 
\ref{fig:pH0_dip_O4}. Then we compute $p(H_0)$ with the same events in the presence of a cosmic dipole with $g=0.01$ and $(l^{\rm dip},b^{\rm dip})=(264^\circ,48^\circ)$ in galactic coordinates, and obtain $H_0=72.26^{+3.89}_{-3.61}~{\rm km}~{\rm s}^{-1}~{\rm Mpc}^{-1}$. Here we choose $g$ an order of magnitude larger than that obtained from the CMB to exaggerate the cosmic dipole effects. However, it shows that in the O4 stage, the cosmic dipole makes little deviations to the combined $p(H_0)$.

\begin{figure}[t]
    \centering
    \includegraphics[width=0.6\linewidth]{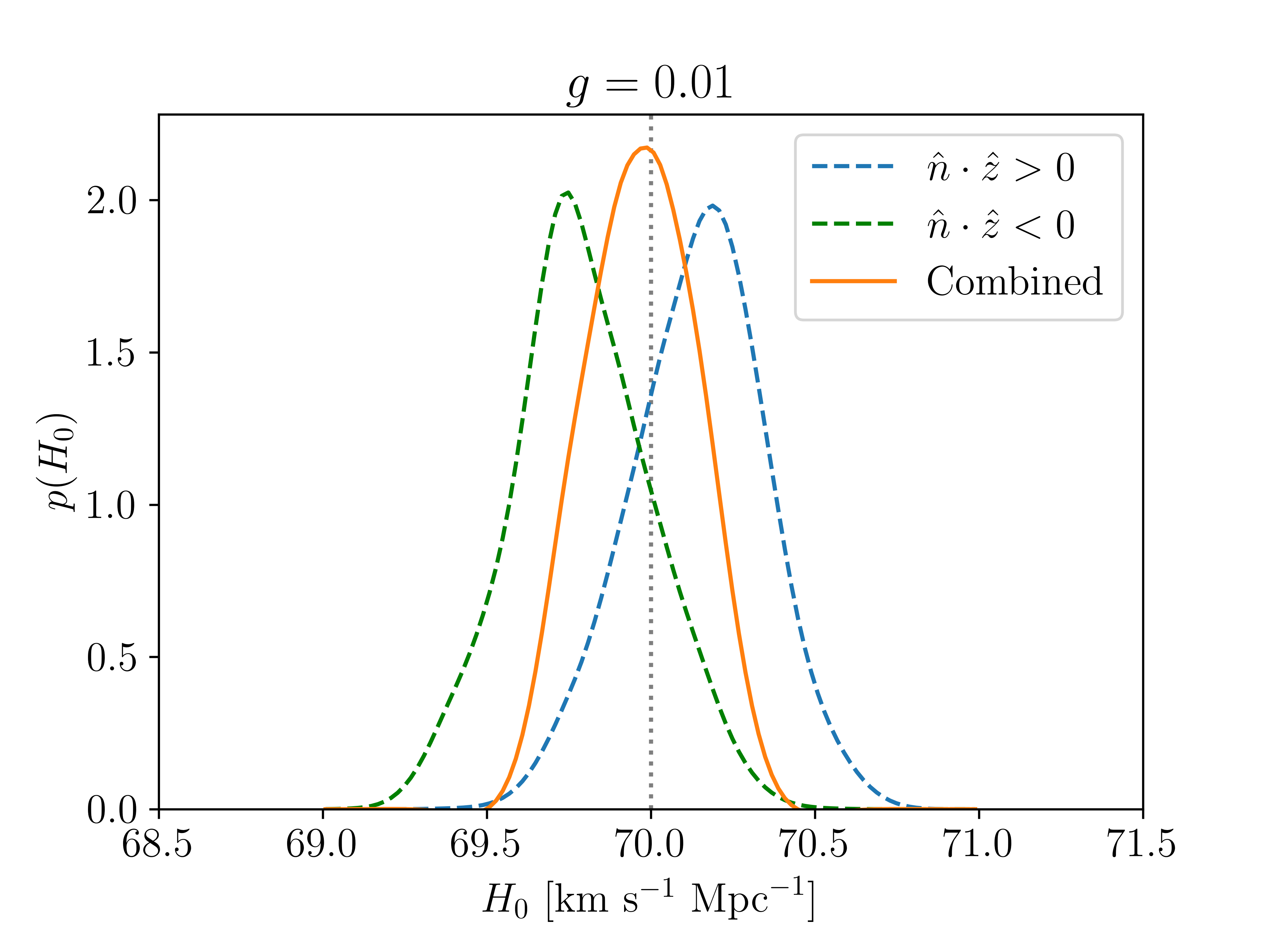}
    \caption{Combined posterior distribution for $H_0$ using mock O5 events with an injected cosmic dipole for $g=0.01$. Orange curve: Combined posterior for all over the sky. Blue and green: Combined posteriors for the two hemispheres in the positive and the negative direction of the dipole.}
    \label{fig:pH0_dip_O5_pos_neg}
\end{figure}

We also examine the same effects in the O5 stage. Using 3073 mock events for 1 year of O5 detections, we compute the combine $p(H_0)$ with the same injected cosmic dipole with $g=0.01$ as in the O4 test. To approximate the results for 5 years of O5 detections, we take the $5^{\rm th}$ power of $p(H_0)$ from 1 year detections assuming $p(H_0)$ in each year is similar to it\footnote{we smooth the curve of $p(H_0)$ in Fig. \ref{fig:pH0_dip_O5_pos_neg}, because \texttt{ICAROGW} approximates the integration in the numerator of eq.~\eqref{eq:bayesian_likelihood} by summing the posterior samples, which leads to wiggles in fine grids.}. Therefore we can obtain a more accurate constraint on $H_0$. We find the combine constraint gives $H_0=69.98^{+0.17}_{-0.19}~{\rm km}~{\rm s}^{-1}~{\rm Mpc}^{-1}$, which is consistent with the injected value of $70$. The results demonstrate that the cosmic dipole has little impact on the $H_0$ measurement with O5 data as well. The reason could be that the biases induced by modification in $D_L^{\rm obs}$ due to the cosmic dipole even out by combining posterior of $H_0$ all over the sky, and that the bias induced by the detection number is not large enough for the O5 stage.

In order to validate our statement above, we compute $p(H_0)$ for events located in the two hemispheres in the positive and the negative direction of the cosmic dipole respectively, as shown in Fig.~\ref{fig:pH0_dip_O5_pos_neg}. The line-of-sight direction of an event $\hat{z}(\phi,\theta)$ is determined by the mean of $\phi$ and $\theta$ from posterior samples. We obtain 1482 events in the hemisphere for $\hat{n}\cdot\hat{z}>0$, and 1591 events for $\hat{n}\cdot\hat{z}<0$. The reason that there are more events for $\hat{n}\cdot\hat{z}<0$ is that the SNRs increase for $D_L^{\rm obs}$ scaled by $[1+g(\hat{n}\cdot\hat{z})]$, so that more events pass the SNR threshold and vice versa. As a result, we find in Fig.~\ref{fig:pH0_dip_O5_pos_neg} that $p(H_0)$ for the two hemispheres peak on different sides of the injected $H_0$ value. The reason is that the likelihood depends on the redshift prior $p_{\rm rate}(z|\Lambda) {\rm d}V_c/{\rm d}z(\Lambda)$, where the growing rate of ${\rm d}V_c/{\rm d}z$ increases with redshift at the low redshift range (LVK detection range). Since higher $D_L^{\rm obs}$ scaled by $[1+g(\hat{n}\cdot\hat{z})]$ can be converted to higher redshift, and the redshift prior has a larger value and a steeper slope at higher redshift, the integration over redshift in eq.~\eqref{eq:bayesian_likelihood} gives larger results compared to $D_L^0$. Therefore the slope of a single $p(H_0)$ becomes larger, and the peak of combined $p(H_0)$ shifts to higher $H_0$ and vice versa. The deviation between the mean values of $H_0$ for the two hemispheres is nearly $1\sigma$ uncertainty. 

One may be able to constrain the cosmic dipole from the deviation between $H_0$ measured in the two hemispheres. However, more careful analysis is required for the dark siren method in different hemispheres, as the selection effects may be different for different line-of-sight directions in the presence of the cosmic dipole. With more detected events by the next-generation detectors, the deviation between $p(H_0)$ in different hemispheres will be larger due to smaller uncertainty for $H_0$. But it is also worth noted that by fixing the population and merger rate model as we did in this work, there would be biases in $p(H_0)$ from real GW data if assuming the wrong models. The safer approach to measure $H_0$ is to jointly measure population and merger rate model parameters, which will weaken the constraint on $H_0$, so that the biases due to the cosmic dipole will become less significant. Moreover, for a dipole at the CMB level, such effects will further reduce.



\section{Measurements of the cosmic dipole}
\label{sec:measurements}

In this section we will introduce the method to jointly measure $H_0$ and the cosmic dipole using golden dark sirens. We simulate mock golden dark sirens in different ground-based detector networks in the future and forecast constraints on $H_0$ and the cosmic dipole with this method.

\subsection{Golden dark sirens}

One of the difficulties in GW cosmology is to obtain the redshifts of GW events, which usually relies on redshifts of their host galaxies. The localization of GWs is very limited nowadays, which could cover many galaxies in the localization area of an event, so that its host galaxy is hard to identify. One exception is bright sirens, whose host galaxies can be pinned down by EM counterparts. But there is also another possible way. From galaxy surveys, an empirical relation between the number density and luminosity of galaxies is summarized as the Schechter function \cite{1976ApJ...203..297S}. The number density of galaxies is lower for galaxies with higher luminosity. With the assumption that galaxies with higher luminosity have higher possibility to host CBCs because of larger stellar masses, one can identify the GW host galaxy in a very small localization area where there is only one brightest galaxy. Based on the Schechter function, it is found in ref. \cite{Singer:2016eax} that for redshift up to $0.1$, the areal density of galaxies at distance $r$ with luminosity greater than $L$ is approximated by
\begin{equation}
    N_{\rm gal} \approx 0.28~{\rm deg}^{-2} \left(\frac{r}{100 {\rm Mpc}} \right)^3 \left(\frac{\phi^*}{4\times10^{-3}{\rm Mpc}^{-3}} \right) \Gamma\left(\alpha+1 ,\frac{L}{L^*}\right),
    \label{eq:ngal_deg2}
\end{equation}
where we adopt the B-band Schechter function parameters $\alpha=-1.25$, $\phi^*=1.2\times10^{-2}~h^3~{\rm Mpc}^{-3}$, and $L^*=1.2\times10^{10}~h^{-2}~L_\odot$ \cite{Longair:2008gba}. Under the assumption of our cosmological parameters $H_0=70~{\rm km}~{\rm s}^{-1}~{\rm Mpc}^{-1}$ and $\Omega_{m0}=0.3$, the luminosity distance at $z=0.1$ is approximately $460$ Mpc. From eq. \eqref{eq:ngal_deg2}, statistically only one galaxy with $L>L^*$ can be found within a sky area of $\sim 0.06~{\rm deg}^2$ up to $z=0.1$. Therefore, for a GW event detected within $\sim 0.06~{\rm deg}^2$ and $z\leq0.1$, the only brightest galaxies in this area has the highest probability to be its host galaxy, from which its redshift can be determined. Such an event is called a ``golden" dark siren, which offers a powerful tool for measuring $H_0$ \cite{Borhanian:2020vyr,Gupta:2022fwd,Chen:2024gdn}. In this work we will explore its potential in probing the cosmic dipole.

We consider the detections of golden dark sirens with three different detector network configurations, namely 1ET+2A\#s, 1ET+1CE+1A\#, and 1ET+2CEs. In the first network, both LIGO detectors have A\# sensitivity. In the second network, we assume that LIGO Livingston is replaced by CE, and LIGO Hanford has A\# sensitivity. In the last network, both LIGO detectors are replaced by CEs. We again use \texttt{GWSim} to generate mock events with detector setups introduced in Sec. \ref{sec:GW_simulation}. We simulate BBHs, NSBHs and BNSs respectively within $z<0.1$ for 10 years of observations. The black hole population model and the CBC merger rate redshift evolution adopt the same parameters as in Table \ref{tab:param}. For the neutron star population, we apply a uniform distribution in the mass range of $[1.0M_\odot,3.0M_\odot]$. In addition, since the merger rates of NSBHs and BNSs have large uncertainties ($7.8$-$140~{\rm Gpc}^{-3}~{\rm yr}^{-1}$ for NSBHs and $10$-$1700~{\rm Gpc}^{-3}~{\rm yr}^{-1}$ for BNSs inferred from GWTC-3 \cite{KAGRA:2021duu}), we assume the same merger rate for BBHs, NSBHs and BNSs as a reasonable guess. 

Since these events locate at very low redshifts, they have very high SNRs. In this scenario, the best-fit parameter values for GW signals would slightly deviate from the true values due to different noise realization, but they follow Gaussian distribution with the mean at the true values. In this case, the uncertainties of the parameters can be forecasted by the Fisher matrix $\boldsymbol{\Gamma}$ as \cite{Poisson:1995ef}
\begin{equation}
	p(\boldsymbol{\theta_p}) \propto \exp\left(-\frac{1}{2}\Gamma_{ij}\Delta\theta_{p,i}\Delta\theta_{p,j}\right).
\end{equation}
The parameter uncertainties are obtained from the covariance matrix $\Delta\theta_{p,i}=\sqrt{\Sigma_{ii}}$, which is the inverse of the Fisher matrix $\Sigma_{ij}=(\boldsymbol{\Gamma}^{-1})_{ij}$. We compute the uncertainties for the GW parameter set $\{m_1^{\rm obs},m_2^{\rm obs},D_L^{\rm obs},\iota,\phi,\theta\}$. Then we select golden dark sirens whose $90\%$ confidence sky area is within $0.06~{\rm deg}^2$. We compute the $90\%$ confidence sky area approximately by multiplying the $2\sigma$ uncertainty ranges of $\phi$ and $\theta$, since the $2\sigma$ uncertainty range cover the $95\%$ confidence interval. For each detector network, the number of each type of event and the total number of detected golden dark sirens are listed in Table \ref{tab:event_number}. We also check the number of golden dark sirens that LVK can detect, where the two LIGO detectors use A\# sensitivity, and Virgo and KAGRA use their designed sensitivities. However only one golden dark siren is detected, which cannot provide a constraint on the cosmic dipole.
\begin{table}[t]
    \centering
    \begin{tabular}{|c|c|c|c|c|}
    \hline
        \multirow{2}{*}{\makecell{Detector\\network}} & \multicolumn{4}{|c|}{Golden dark siren number} \\
        \cline{2-5}
         & BBH & NSBH & BNS & Total \\
        \hline
        1ET+2A\#s & 13 & 0 & 0 & 13 \\
        \hline
        1ET+1CE+1A\# & 31 & 2 & 2 & 35 \\
        \hline
        1ET+2CEs & 33 & 3 & 3 & 39  \\
    \hline
    \end{tabular}
    \caption{Detected number of golden dark sirens for different detector networks for 10-year observations.}
    \label{tab:event_number}
\end{table}

\begin{figure*}[t]
    \centering
    \includegraphics[width=\textwidth]{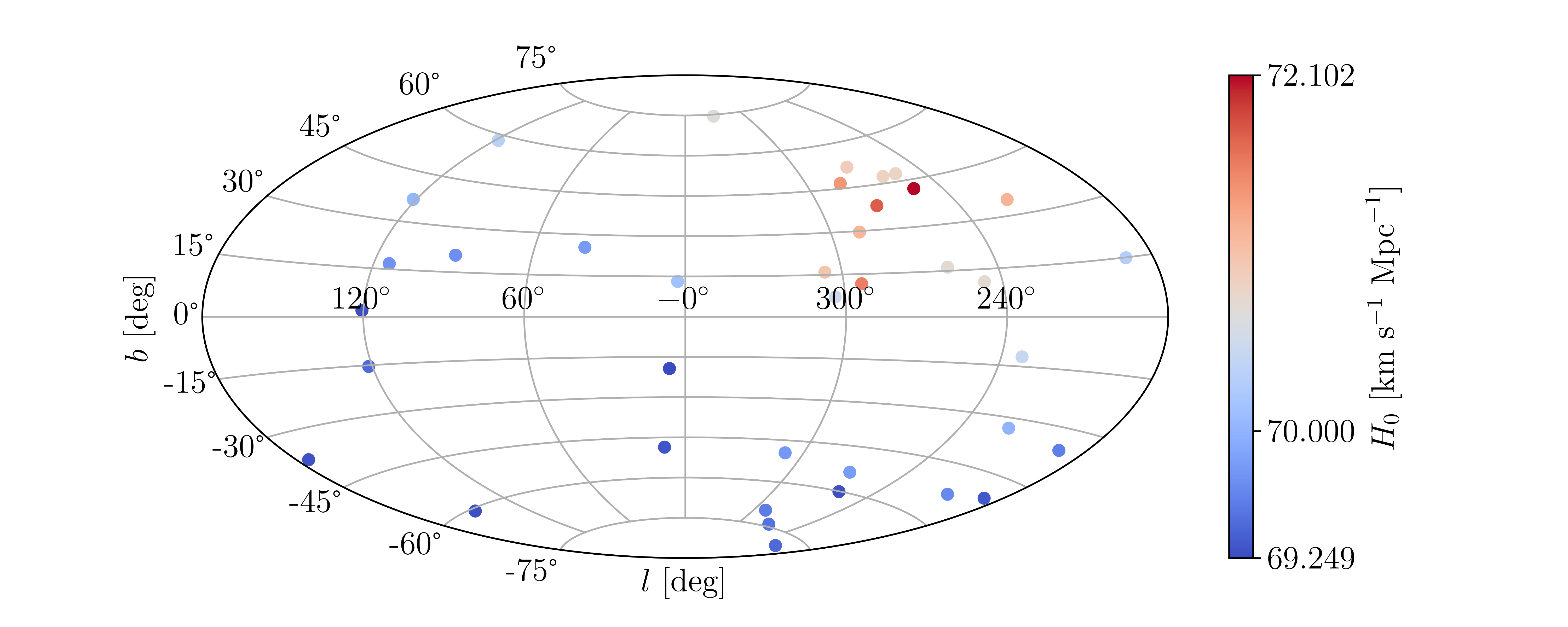}
    \caption{Skymap showing the differences in $H_0$ measured from each golden dark sirens detected by the 1ET+2CEs network when not accounting for the cosmic dipole effects. The injected $H_0=70~{\rm km}~{\rm s}^{-1}~{\rm Mpc}^{-1}$. The injected dipole has $g=0.001$ and $(l^{\rm dip},b^{\rm dip})=(264^\circ,48^\circ)$.}
    \label{fig:skymap_H0_XG}
\end{figure*}

We show in Fig. \ref{fig:skymap_H0_XG} the locations of simulated golden dark sirens detected by the 1ET+2CEs network. In the figure we also demonstrate the biases in the $H_0$ measurement using $D_L^{\rm obs}$ and $z^{\rm obs}$ for these golden dark sirens if not accounting for the cosmic dipole effects. For an injected cosmic dipole magnitude $g=0.001$ and the CMB dipole direction $(l^{\rm dip},b^{\rm dip})=(264^\circ,48^\circ)$, the obtained maximum $H_0$ value is around $72.1~{\rm km}~{\rm s}^{-1}~{\rm Mpc}^{-1}$, which has a bias of around $3\%$ compared to the injected value of $70~{\rm km}~{\rm s}^{-1}~{\rm Mpc}^{-1}$. The minimum $H_0$ value is lower than the injected value by $\sim 1\%$, but since no mock event is located near the negative cosmic dipole direction, the real minimum $H_0$ value could have a larger bias similar to the maximum value. Given a sub-percent level of uncertainty in $H_0$ measurement with golden dark sirens detected by an ET+CE network discussed in \cite{Chen:2024gdn}, a bias of $3\%$ could be fatal for such measurement.

\subsection{MCMC sampling}

After we obtain mock golden dark siren data, we inject the cosmic dipole effects with $g=0.001$ and $(l^{\rm dip},b^{\rm dip})=(264^\circ,48^\circ)$ by modifying the observed quantities $m_1^{\rm obs}, m_2^{\rm obs}, D_{L}^{\rm obs}$ for GWs and $z^{\rm obs}$ for host galaxies of golden dark sirens. Here we choose $g$ at the CMB level as the lower observed value to demonstrate the detectability of the golden dark sirens. We then measure the cosmic dipole jointly with $H_0$ by $\chi^2$ minimization using the Markov Chain Monte Carlo (MCMC) sampling technique. The log-likelihood for $\chi^2$ statistics is given by
\begin{equation}
    \log{\cal L} = -\frac{\chi^2}{2},
\end{equation}
where we build
\begin{equation}
    \chi^2 = \sum_i^{N_{\rm obs}} \left\{\frac{D_{L,i}^{\rm obs}-D_L^0(z^0(z_i^{\rm obs},g,\hat{n}),H_0)[1+g(\hat{n}\cdot\hat{z}_i)]}{\Delta D_{L,i}^{\rm obs}}\right\}^2.
\end{equation}
Here $\Delta D_{L,i}^{\rm obs}$ is the uncertainty from Fisher matrix analysis. We restore $D_L^0$ in the CMB rest frame with $H_0$ and redshift in the CMB rest frame $z^0$, which is obtained by
\begin{equation}
    z^0(z_i^{\rm obs},g,\hat{n}) = \frac{1+z_i^{\rm obs}}{1+g(\hat{n}\cdot\hat{z}_i)}-1.
    \label{eq:z0_convert}
\end{equation}
We then perform MCMC sampling with the \texttt{emcee} package \cite{emcee} to find the best-fit values for $\{H_0, g, l^{\rm dip}, b^{\rm dip}\}$.

\section{Results}
\label{sec:results}

\begin{figure}[t]
    \centering
    \includegraphics[width=0.8\linewidth]{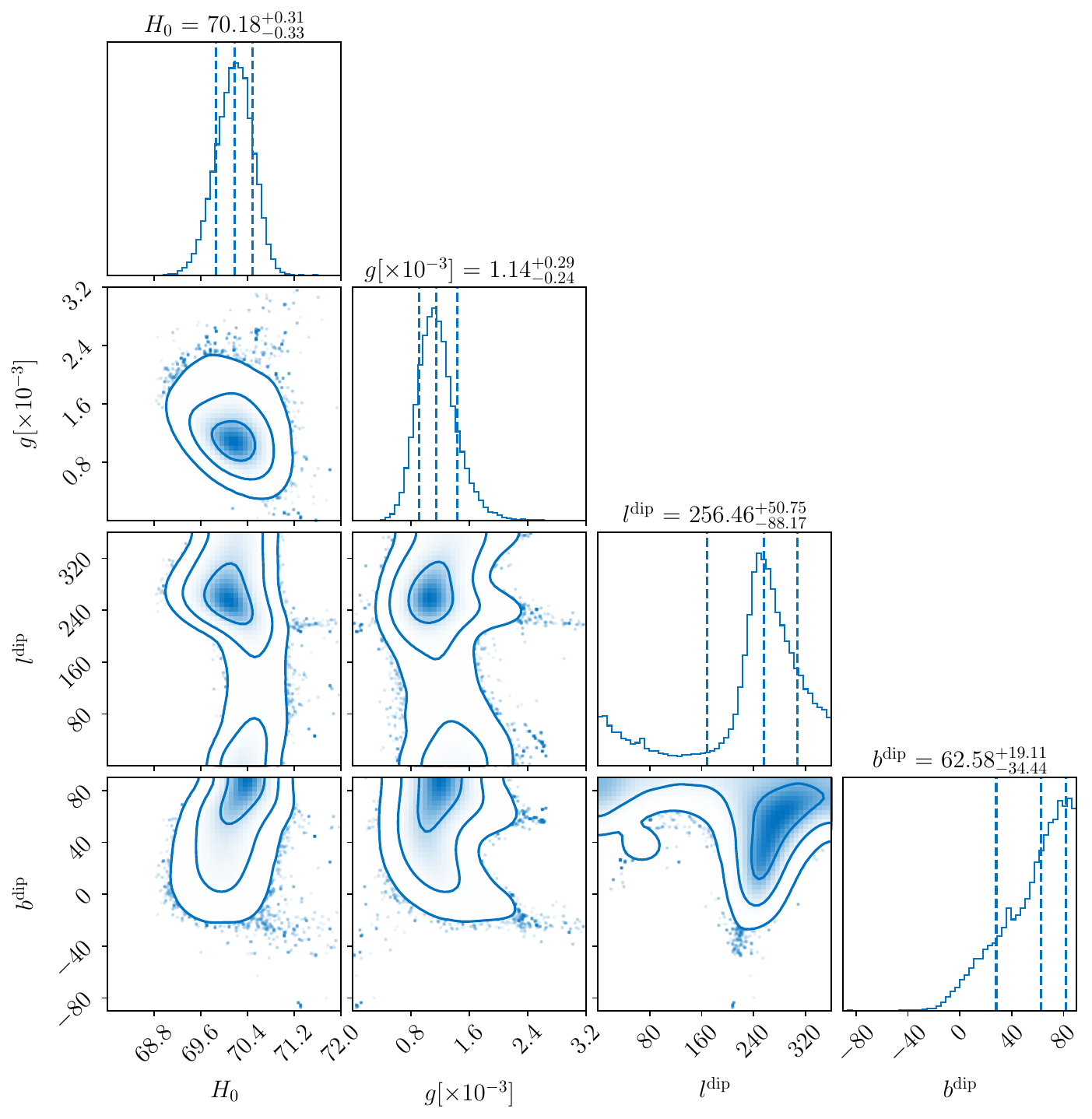}
    \caption{Joint posterior for $H_0$ and the dipole parameters form 13 golden dark sirens by 1ET+2A\# network, with an injected value of $g=10^{-3}$.}
    \label{fig:emcee_dip_g0p001_H0_corner_ET_AS}
\end{figure}

\begin{figure}[t]
    \centering
    \includegraphics[width=0.8\linewidth]{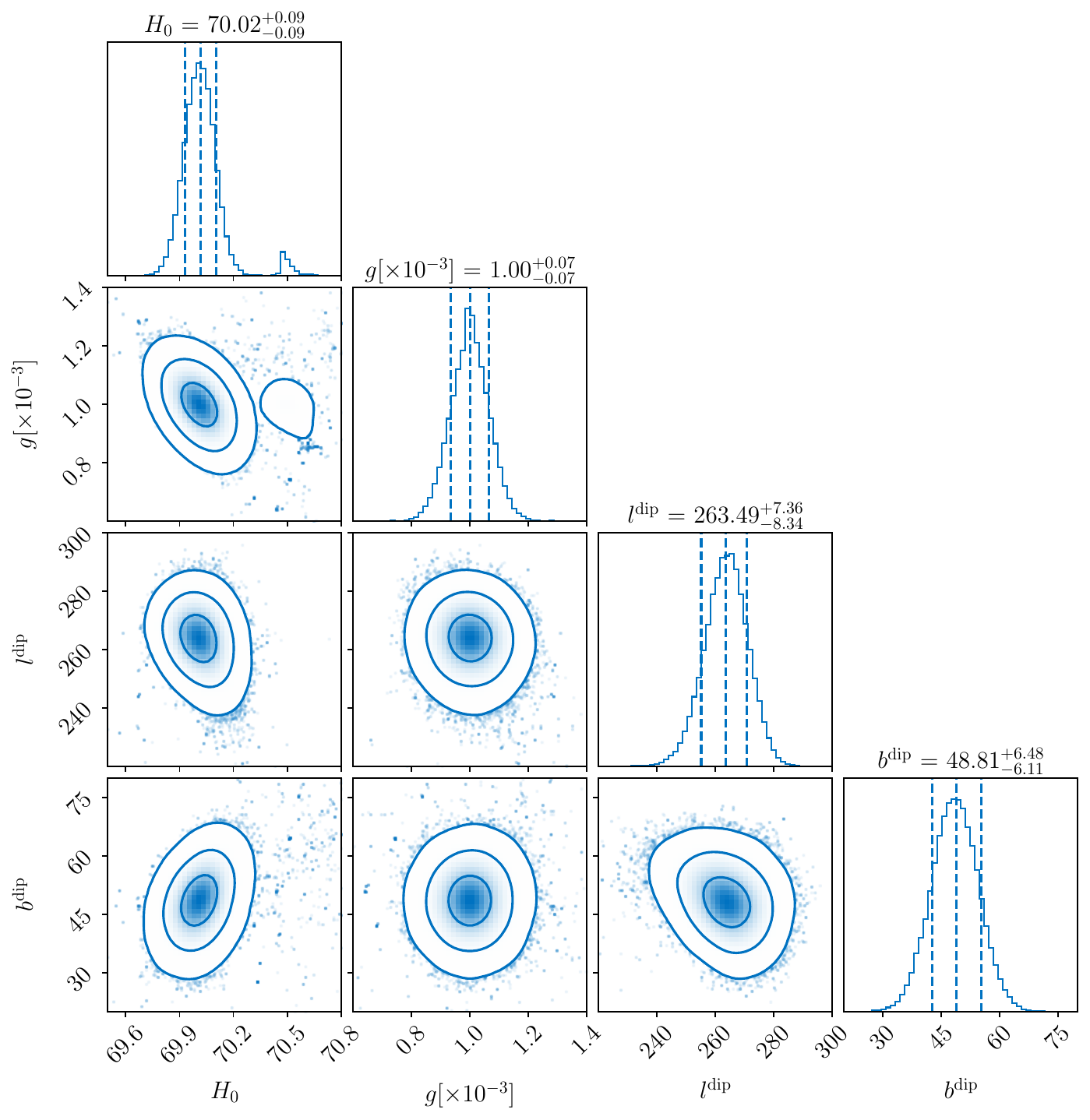}
    \caption{Joint posterior for $H_0$ and the dipole parameters form 35 golden dark sirens by 1ET+1CE+1A\# network, with an injected value of $g=10^{-3}$.}
    \label{fig:emcee_dip_g0p001_H0_corner_XG_AS}
\end{figure}

\begin{figure}[t]
    \centering
    \includegraphics[width=0.8\linewidth]{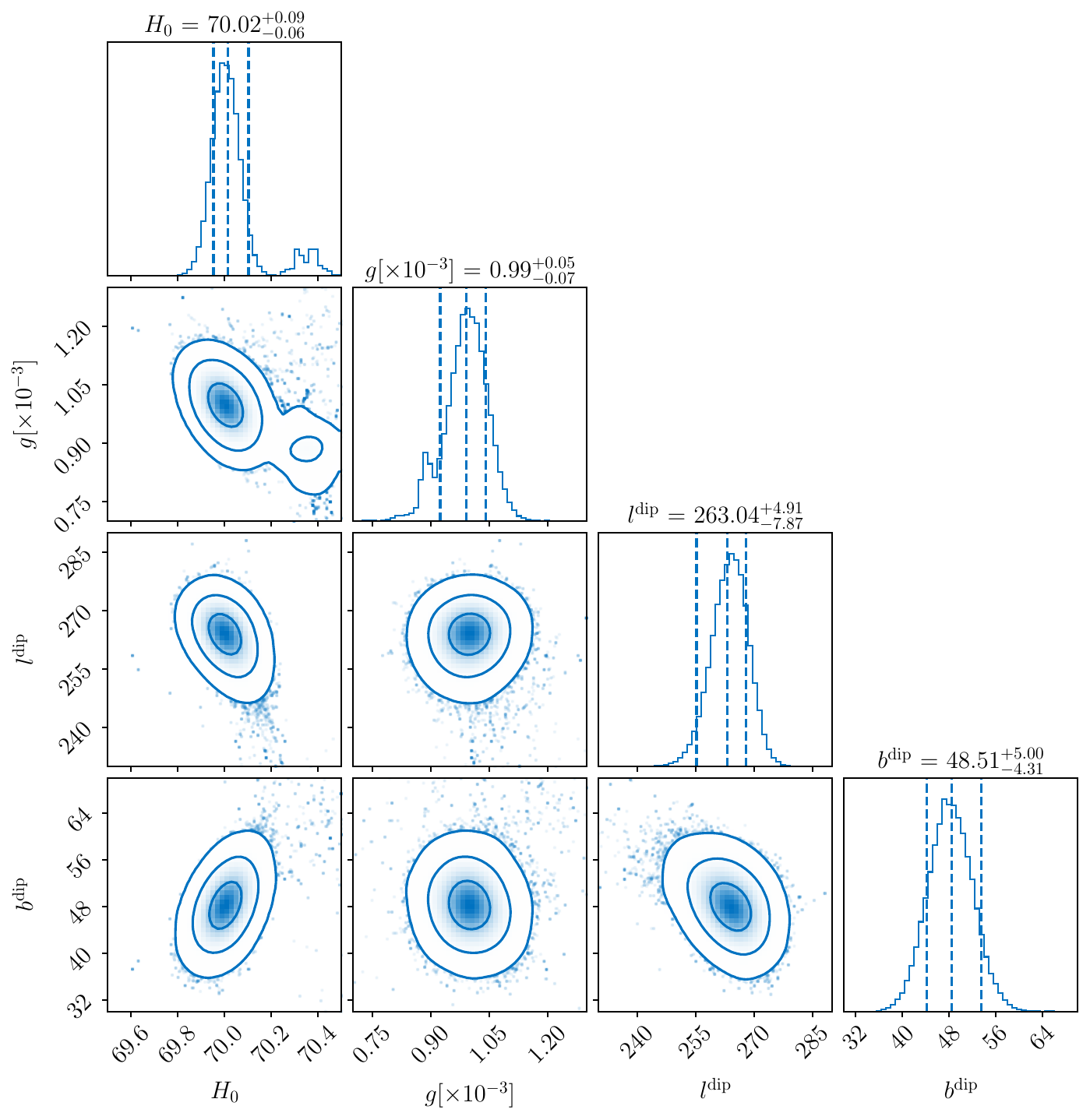}
    \caption{Joint posterior for $H_0$ and the dipole parameters form 39 golden dark sirens by 1ET+2CE network, with an injected value of $g=10^{-3}$.}
    \label{fig:emcee_dip_g0p001_H0_corner}
\end{figure}

We show the joint posteriors of $H_0$ and the cosmic dipole parameters from MCMC sampling in the corner plots in this section. For the pessimistic case that only one ET is built, so that the detector network consists of ET and two LIGO detectors with A\#s sensitivity, the result is presented in Fig. \ref{fig:emcee_dip_g0p001_H0_corner_ET_AS}. Although $H_0$ can be constrained at around $0.46\%$ uncertainty, the detected golden dark siren number is too few to give a tight constraint on the cosmic dipole. The $1\sigma$ uncertainty for the dipole magnitude is around $27\%$ for an injected value of $g=0.001$, and the uncertainty area of the dipole direction is over $800~{\rm deg}^2$. But with at least one CE, the constraints on $g$ and $(l^{\rm dip}, b^{\rm dip})$ become much tighter as shown in Fig \ref{fig:emcee_dip_g0p001_H0_corner_XG_AS} and \ref{fig:emcee_dip_g0p001_H0_corner} because of about two times increase in the number of golden dark sirens. The $1\sigma$ uncertainties of measurements for each network are listed in Table \ref{tab:g_emcee}.

It is shown that by combining ET with one CE and one LIGO detector, 35 golden dark sirens can be detected, which can thus reduce the uncertainties of all parameters by around $3/4$. Moreover, in the most optimistic case where both LIGO detectors are replaced by CEs, the constraints are further improved slightly, since the golden dark siren number increases by 4. The cosmic dipole magnitude $g$ is constrained down to $6\%$ uncertainty. Meanwhile, by multiplying the $1\sigma$ uncertainty range of $l^{\rm dip}$ and $b^{\rm dip}$, the dipole direction can be constrained within $120~{\rm deg}^2$. This constraint is not as accurate as the CMB constraint, but it is slightly better than the constraints obtained by radio source number counting. 

On the other hand, our constraint is comparable to the constraints forecasted by bright sirens with the next-generation detector network for 10-year observations in \cite{Cousins:2024bhk}. Although the golden dark sirens we found are much fewer than bright sirens forecasted in their work, they are mostly BBHs located at low redshifts, and thus have much larger SNRs. As a result, their observed luminosity distances are measured more precisely, which leads to better constraints on $H_0$ and the cosmic dipole. A second reason is that we convert the observed redshift into the rest-frame redshift as in eq. \eqref{eq:z0_convert}, and then construct the rest-frame luminosity distance from it, which is a step neglected in \cite{Cousins:2024bhk}. This conversion also provides constraint on the cosmic dipole.

\begin{figure}[t]
    \centering
    \includegraphics[width=0.6\textwidth]{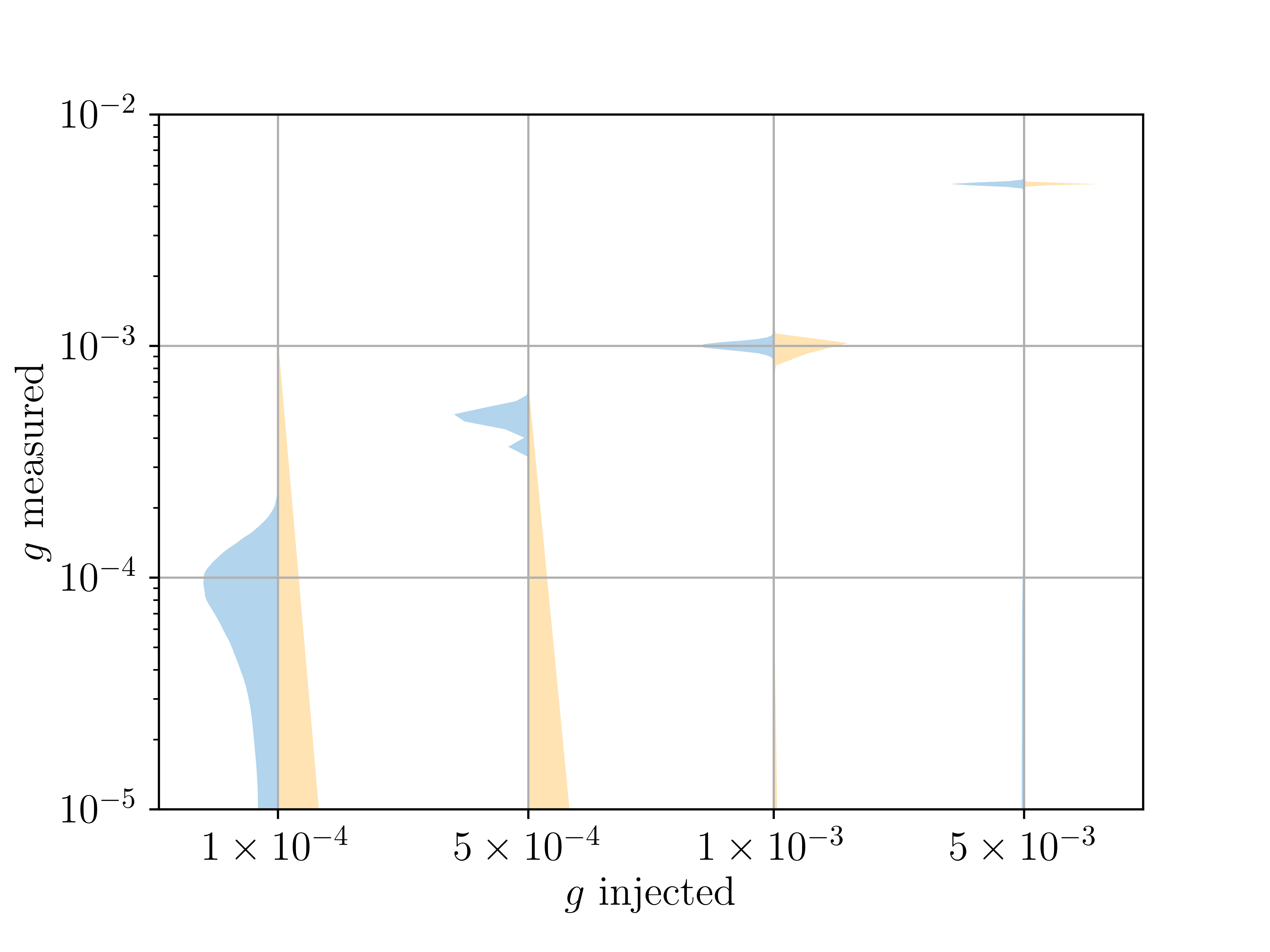}
    \caption{Violin plot showing the marginalized posteriors of $g$ by MCMC sampling using mock golden dark sirens detected by the 1ET+2CEs network for each injected value of $g$. Orange violins show the results for joint estimation with $H_0$, and blue violins show the results for fixing $H_0$.}
    \label{fig:g_violin}
\end{figure}

\begin{figure}
\begin{subfigure}
    \centering
    \includegraphics[width=0.5\linewidth]{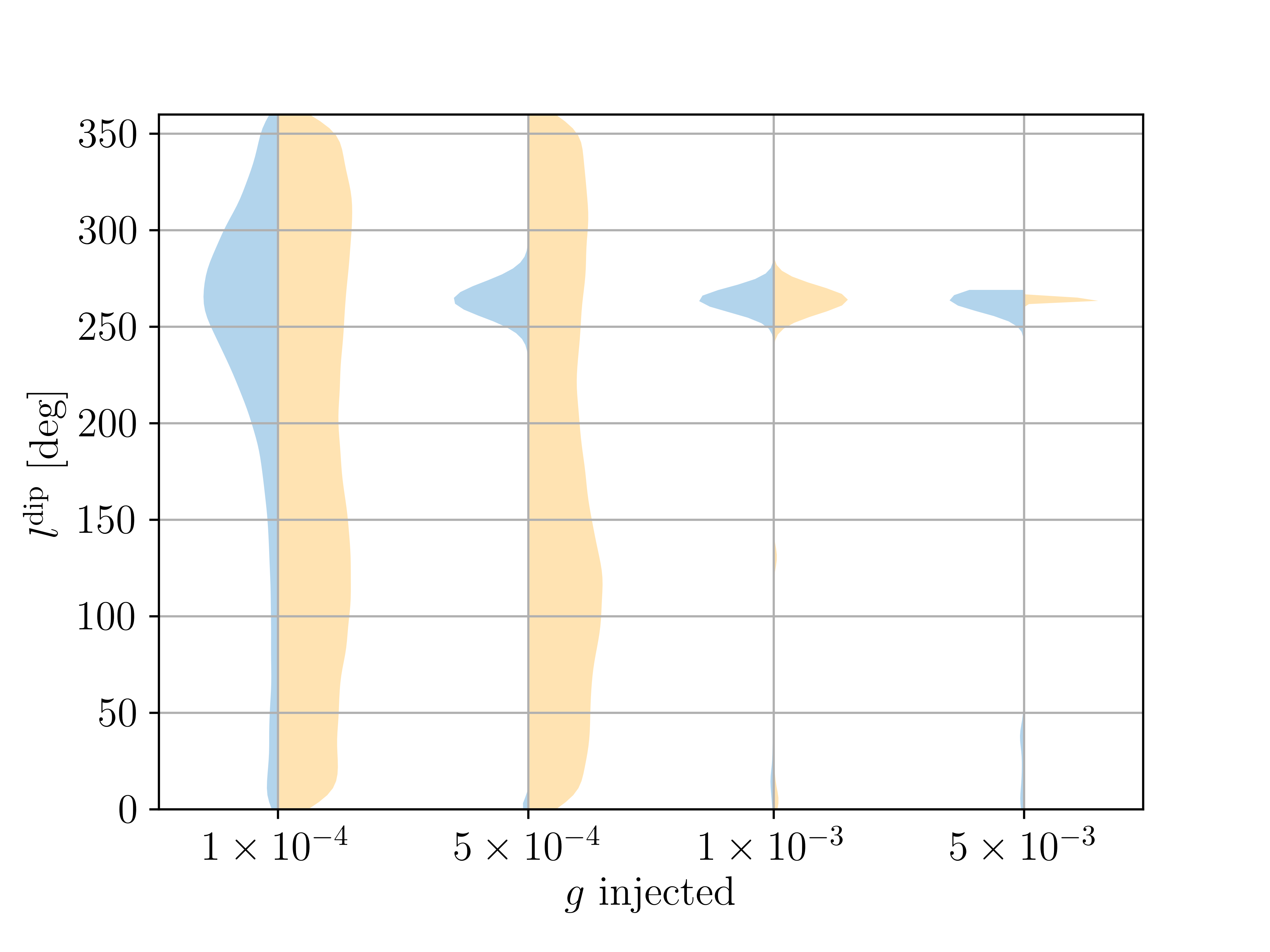}
\end{subfigure}
\begin{subfigure}
    \centering
    \includegraphics[width=0.5\linewidth]{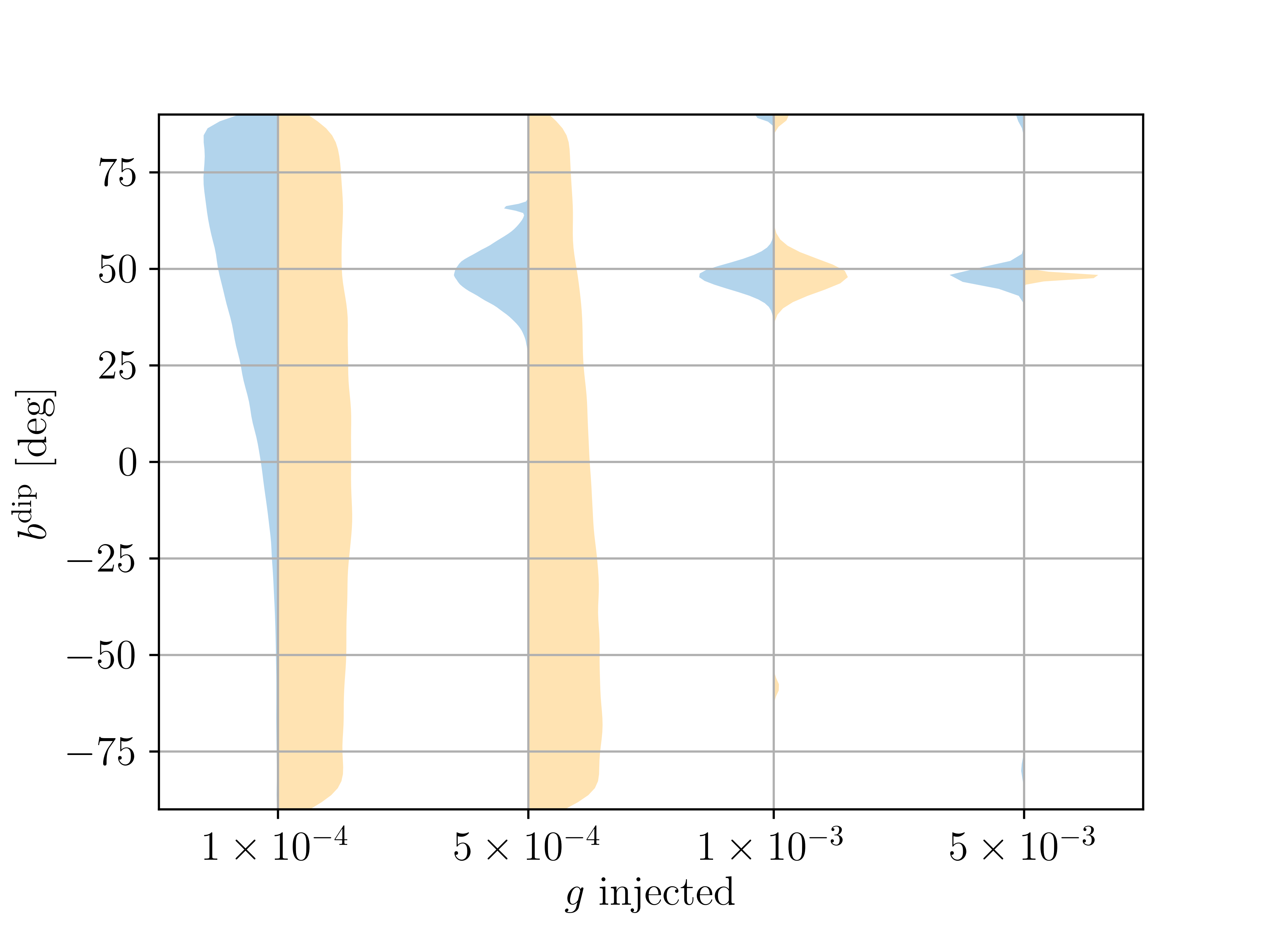}
\end{subfigure}
    \caption{Violin plot showing the marginalized posteriors of $l^{\rm dip}$ (left panel) and $b^{\rm dip}$ (right panel) by MCMC sampling using mock golden dark sirens detected by the 1ET+2CEs network for each injected value of $g$. Orange violins show the results for joint estimation with $H_0$, and blue violins show the results for fixing $H_0$.}
    \label{fig:dec_violin}
\end{figure}

We also investigate the constraints on the cosmic dipole for different injected values of the dipole magnitude $g$, but keeping the dipole direction unchanged. We use the same set of simulated golden dark sirens detected by the most optimistic network 1ET+2CEs, but vary the value of $g$ when modifying $m_1^{\rm obs}$, $m_2^{\rm obs}$, $D_L^{\rm obs}$ and $z^{\rm obs}$. We then perform Fisher matrix analyses for golden dark siren datasets with different injected values of $g$.
We run MCMC sampling to measure $g$, $l^{\rm dip}$ and $b^{\rm dip}$ with and without joint estimation on $H_0$ respectively. The marginalized posteriors in both cases are plotted on both sides of the violin plots in Fig. \ref{fig:g_violin} and \ref{fig:dec_violin} for $g$ and the direction (orange violins for joint estimation with $H_0$, and blue violins for without it). 

We find that when jointly measured with $H_0$, the dipole magnitude and direction can be well constrained for $g$ in the order of $10^{-3}$. However, for $g$ in the order of $10^{-4}$, we cannot obtain a meaningful constraint on the dipole parameters. On the other hand, when measuring the cosmic dipole with fixed $H_0$, the constraints for $g$ in the order of $10^{-3}$ are similar to those jointly measured with $H_0$. But we can still obtain strong constraints on $g$ and the dipole direction for $g$ in the order of $10^{-4}$. The results show that joint estimation with $H_0$ lowers the constrainability of the dipole magnitude and direction. Although the joint posterior corner plot in Fig. \ref{fig:emcee_dip_g0p001_H0_corner} shows little degeneracy between $H_0$ and the cosmic dipole parameters, including more parameters in the sampling process could lead to such downgrade in the cosmic dipole constraints.
\begin{table}[t]
    \centering
    \begin{tabular}{|c|c|c|c|c|}
    \hline
        \multirow{2}{*}{Detector network} & \multicolumn{4}{|c|}{$1\sigma$ uncertainty} \\
        \cline{2-5}
         & $H_0$ & $g$ & $l^{\rm dip}$ & $b^{\rm dip}$ \\
        \hline
        1ET+2A\#s & $0.46\%$ & $27\%$ & - & - \\
        \hline
        1ET+1CE+A\# & $0.13\%$ & $7\%$ & $8^\circ$ & $6^\circ$  \\
        \hline
        1ET+2CEs & $0.11\%$ & $6\%$ & $6^\circ$ & $5^\circ$ \\
    \hline
    \end{tabular}
    \caption{$1\sigma$ uncertainty for parameter measurements by MCMC sampling with forecasted golden dark sirens by different detector networks. For the 1ET+2A\#s network, $l^{\rm dip}$ and $b^{\rm dip}$ have non-Gaussian distribution, so the 1$\sigma$ errors are not obtained.}
    \label{tab:g_emcee}
\end{table}

Finally, we compare our method with the GW number counting method in \cite{Mastrogiovanni:2022nya,Grimm:2023tfl}. The number counting method can measure the comic dipole magnitude at the CMB level $1.2\times10^{-3}$ with an error lower than $20\%$ using $10^7$ events \cite{Grimm:2023tfl}, which also requires a few years of observations by ET and CE. The golden dark sirens observed in 10 years can yield a constraint on the dipole magnitude at the CMB level with a much lower error, which is under $10\%$ with ET and CE. Even if the observation time cannot reach 10 years, a comparable constraint to the number counting method is expected to be obtained by golden dark sirens. More importantly, our method constrains the cosmic dipole magnitude and direction jointly with $H_0$, which is not contained in other works. Nevertheless, by combinging the number counting method and the golden dark siren method, a stronger constraint on the cosmic dipole can be achieved.

\section{Conclusions}
\label{sec:conclusions}

In the first part of this work, we investigated the cosmic dipole effects in $H_0$ measurement with the dark siren method. Using simulated BBHs detected in the LVK O4 and O5 observations, we found that the cosmic dipole effects won't cause significant biases in $p(H_0)$ using the population method with a fixed population model.  The reason could be that the deviation in the number of GW detections for LVK induced by the cosmic dipole is too small, which is in the order of $10$ for 5 years of O5 observations. When combining $p(H_0)$ for events all over the sky, the cosmic dipole effects on $D_L^{\rm obs}$ cancel out for events in hemispheres on both sides of the cosmic dipole direction. Therefore no significant bias is found in the combined $p(H_0)$.

However, the cosmic dipole effects could become significant in the era of next-generation GW detectors. Since ET and CE are expected to detect $10^5$-$10^6$ events per year, the deviation in the detection number due to the cosmic dipole could reach an order of $10^3$-$10^4$, which could bring significant biases in $H_0$ measurement with the dark siren method. The current dark siren tools are too computational expensive to handle the large number of events by next-generation detectors, so the forecast of the cosmic dipole effects with ET and CE observations is not easy to perform at the moment. In addition, the selection effects on sky localization should also be taken into account for the future dark siren technique, because the detection probability could change due to the cosmic dipole effects on $D_L^{\rm obs}$. Moreover, when incorporating redshift information from galaxy catalogues in the dark siren method, $H_0$ measurement could be more accurate, so the cosmic dipole effects could be more significant as well. We leave further investigation on these topics to future's works.

In the second part of our work, we proposed to use golden dark sirens detected by the next-generation detector networks to measure the cosmic dipole jointly with $H_0$. We found that for dipole magnitude $g=0.001$, a network consisting of three detectors, where at least two next-generation detectors like ET and CE are included, can yield a tight constrain on the dipole magnitude and the direction. With the most optimistic network 1ET+2CEs, $g$ can be constrained down to $6\%$, and the dipole direction can be constrained within $120~{\rm deg}^2$. In addition, we also found that when $g$ is at the order of $10^{-4}$, it cannot be constrained using golden dark sirens with joint estimation with $H_0$. However, when fixing $H_0$ in the measurement, moderate constraints can still be obtained for the dipole magnitude and the direction.

The precise measurement of the cosmic dipole by golden dark sirens would serve as an independent probe to the cosmic dipole in the future. It could provide hints to solve the cosmic dipole tension between measurements by the CMB and source number counting in the local universe. Moreover, the joint measurement of $H_0$ could also play a significant role in tackling the Hubble tension between the early and the late universe. Since dark siren events are expected to have astrophysical origins, their distribution is expected to follow galaxy distribution, so the estimation of the cosmic dipole from GW number counting is likely to give similar results to galaxy number counting. However, the estimation from golden dark sirens does not rely on number counting. If it yields similar results to the number counting method, the cosmic dipole tension is likely due to unknown cosmology between the early and the late universe. But if golden dark siren measurement favours the CMB value, it could point out that systematics likely exist in the number counting method.

It should be noted that our measurement of the cosmic dipole depends on the assumption that the host galaxies of the golden dark sirens are correctly identified. The probability for the brightest galaxy within the localization area of a golden dark siren to be its host galaxy is not quantitatively calculated. It could be possible that the host galaxy of a GW source is not the brightest galaxy. Therefore quantitative calculation of the host galaxy probability are needed in future's works to enhance the accuracy of the golden dark siren measurement. On the other hand, if GWs from BNSs and NSBHs are accompanied by EM counterparts such as kilonova, their host galaxies can also be pinned down. By including bright sirens in the joint analysis with golden dark sirens, the precision of cosmic dipole measurement can be enhanced due to the increased event number. However, the detection rate of such bright sirens has a large uncertainty from the current data, which makes it difficult to perform an accurate forecast. In addition, the peculiar velocity of host galaxies of GW events may cause systematics in the results (see \cite{Nishizawa:2010xx,Nimonkar:2023pyt} for systematics in measuring $H_0$), which should be taken into account for a more precise measurement in future works. Furthermore, a thorough theoretical framework of kinematic effects on GWs derived in \cite{Cusin:2024git} is essential for a more comprehensive study.

In summary, although the cosmic dipole has little impact to the dark siren cosmology with LVK detections, more careful consideration is needed for the dark siren method in the era of next-generation detectors. With at least one ET and one CE, a strong constraint on the cosmic dipole can be obtained with the golden dark sirens, which can provide an independent measurement of the cosmic dipole and shed light on the cosmic dipole tension.

\section*{Acknowledgments}

The author is grateful to Jun Zhang, Johannes Noller and Martin Hendry for useful comments on this work. A C is supported by the National Natural Science Foundation of China (NSFC) under Grant No. E414660101 and 12147103.
This material is based upon work supported by NSF's LIGO Laboratory which is a major facility fully funded by the National Science Foundation.
We are grateful for computational resources provided by the LIGO Laboratory and supported by the National Science Foundation Grants PHY-0757058 and PHY-0823459.
We are also grateful to the High Performance Computing Center (HPCC) of ICTP-AP for performing the numerical computations in this paper.

\appendix


\bibliographystyle{utphys}
\bibliography{biblio}

\providecommand{\href}[2]{#2}\begingroup\raggedright\begin{thebibliography}{10}

\bibitem{Planck:2018vyg}
{\bfseries Planck} Collaboration, N.~Aghanim {\em et~al.}, ``{Planck 2018
  results. VI. Cosmological parameters},''
  \href{http://dx.doi.org/10.1051/0004-6361/201833910}{{\em Astron. Astrophys.}
  {\bfseries 652} (2021) C4}, \href{http://arxiv.org/abs/1807.06209}{{\ttfamily
  arXiv:1807.06209 [astro-ph.CO]}}.

\bibitem{Abdalla:2022yfr}
E.~Abdalla {\em et~al.}, ``{Cosmology intertwined: A review of the particle
  physics, astrophysics, and cosmology associated with the cosmological
  tensions and anomalies},''
  \href{http://dx.doi.org/10.1016/j.jheap.2022.04.002}{{\em JHEAp} {\bfseries
  34} (2022) 49--211}, \href{http://arxiv.org/abs/2203.06142}{{\ttfamily
  arXiv:2203.06142 [astro-ph.CO]}}.

\bibitem{Riess:2019cxk}
A.~G. Riess, S.~Casertano, W.~Yuan, L.~M. Macri, and D.~Scolnic, ``{Large
  Magellanic Cloud Cepheid Standards Provide a 1\% Foundation for the
  Determination of the Hubble Constant and Stronger Evidence for Physics beyond
  $\Lambda$CDM},'' \href{http://dx.doi.org/10.3847/1538-4357/ab1422}{{\em
  Astrophys. J.} {\bfseries 876} no.~1, (2019) 85},
  \href{http://arxiv.org/abs/1903.07603}{{\ttfamily arXiv:1903.07603
  [astro-ph.CO]}}.

\bibitem{Planck:2018nkj}
{\bfseries Planck} Collaboration, N.~Aghanim {\em et~al.}, ``{Planck 2018
  results. I. Overview and the cosmological legacy of Planck},''
  \href{http://dx.doi.org/10.1051/0004-6361/201833880}{{\em Astron. Astrophys.}
  {\bfseries 641} (2020) A1}, \href{http://arxiv.org/abs/1807.06205}{{\ttfamily
  arXiv:1807.06205 [astro-ph.CO]}}.

\bibitem{Kolb:2005da}
E.~W. Kolb, S.~Matarrese, and A.~Riotto, ``{On cosmic acceleration without dark
  energy},'' \href{http://dx.doi.org/10.1088/1367-2630/8/12/322}{{\em New J.
  Phys.} {\bfseries 8} (2006) 322},
  \href{http://arxiv.org/abs/astro-ph/0506534}{{\ttfamily
  arXiv:astro-ph/0506534}}.

\bibitem{Bolejko:2016qku}
K.~Bolejko and M.~Korzy\'nski, ``{Inhomogeneous cosmology and backreaction:
  Current status and future prospects},''
  \href{http://dx.doi.org/10.1142/S0218271817300117}{{\em Int. J. Mod. Phys. D}
  {\bfseries 26} no.~06, (2017) 1730011},
  \href{http://arxiv.org/abs/1612.08222}{{\ttfamily arXiv:1612.08222 [gr-qc]}}.

\bibitem{Secrest:2020has}
N.~J. Secrest, S.~von Hausegger, M.~Rameez, R.~Mohayaee, S.~Sarkar, and
  J.~Colin, ``{A Test of the Cosmological Principle with Quasars},''
  \href{http://dx.doi.org/10.3847/2041-8213/abdd40}{{\em Astrophys. J. Lett.}
  {\bfseries 908} no.~2, (2021) L51},
  \href{http://arxiv.org/abs/2009.14826}{{\ttfamily arXiv:2009.14826
  [astro-ph.CO]}}.

\bibitem{Blake:2002gx}
C.~Blake and J.~Wall, ``{Detection of the velocity dipole in the radio galaxies
  of the nrao vla sky survey},'' \href{http://dx.doi.org/10.1038/416150a}{{\em
  Nature} {\bfseries 416} (2002) 150--152},
  \href{http://arxiv.org/abs/astro-ph/0203385}{{\ttfamily
  arXiv:astro-ph/0203385}}.

\bibitem{2011ApJ...742L..23S}
A.~K. {Singal}, ``{Large Peculiar Motion of the Solar System from the Dipole
  Anisotropy in Sky Brightness due to Distant Radio Sources},''
  \href{http://dx.doi.org/10.1088/2041-8205/742/2/L23}{{\em Astrophys. J.
  Lett.} {\bfseries 742} no.~2, (Dec., 2011) L23},
  \href{http://arxiv.org/abs/1110.6260}{{\ttfamily arXiv:1110.6260
  [astro-ph.CO]}}.

\bibitem{Gibelyou:2012ri}
C.~Gibelyou and D.~Huterer, ``{Dipoles in the Sky},''
  \href{http://dx.doi.org/10.1111/j.1365-2966.2012.22032.x}{{\em Mon. Not. Roy.
  Astron. Soc.} {\bfseries 427} (2012) 1994--2021},
  \href{http://arxiv.org/abs/1205.6476}{{\ttfamily arXiv:1205.6476
  [astro-ph.CO]}}.

\bibitem{Tiwari:2013vff}
P.~Tiwari, R.~Kothari, A.~Naskar, S.~Nadkarni-Ghosh, and P.~Jain, ``{Dipole
  anisotropy in sky brightness and source count distribution in radio NVSS
  data},'' \href{http://dx.doi.org/10.1016/j.astropartphys.2014.06.004}{{\em
  Astropart. Phys.} {\bfseries 61} (2014) 1--11},
  \href{http://arxiv.org/abs/1307.1947}{{\ttfamily arXiv:1307.1947
  [astro-ph.CO]}}.

\bibitem{Fernandez-Cobos:2013fda}
R.~Fern\'andez-Cobos, P.~Vielva, D.~Pietrobon, A.~Balbi,
  E.~Mart\'\i{}nez-Gonz\'alez, and R.~B. Barreiro, ``{Searching for a dipole
  modulation in the large-scale structure of the Universe},''
  \href{http://dx.doi.org/10.1093/mnras/stu749}{{\em Mon. Not. Roy. Astron.
  Soc.} {\bfseries 441} no.~3, (2014) 2392--2397},
  \href{http://arxiv.org/abs/1312.0275}{{\ttfamily arXiv:1312.0275
  [astro-ph.CO]}}.

\bibitem{Tiwari:2013ima}
P.~Tiwari and P.~Jain, ``{Dipole Anisotropy in Integrated Linearly Polarized
  Flux Density in NVSS Data},''
  \href{http://dx.doi.org/10.1093/mnras/stu2535}{{\em Mon. Not. Roy. Astron.
  Soc.} {\bfseries 447} (2015) 2658--2670},
  \href{http://arxiv.org/abs/1308.3970}{{\ttfamily arXiv:1308.3970
  [astro-ph.CO]}}.

\bibitem{Bengaly:2017slg}
C.~A.~P. Bengaly, R.~Maartens, and M.~G. Santos, ``{Probing the Cosmological
  Principle in the counts of radio galaxies at different frequencies},''
  \href{http://dx.doi.org/10.1088/1475-7516/2018/04/031}{{\em JCAP} {\bfseries
  04} (2018) 031}, \href{http://arxiv.org/abs/1710.08804}{{\ttfamily
  arXiv:1710.08804 [astro-ph.CO]}}.

\bibitem{Siewert:2020krp}
T.~M. Siewert, M.~Schmidt-Rubart, and D.~J. Schwarz, ``{Cosmic radio dipole:
  Estimators and frequency dependence},''
  \href{http://dx.doi.org/10.1051/0004-6361/202039840}{{\em Astron. Astrophys.}
  {\bfseries 653} (2021) A9}, \href{http://arxiv.org/abs/2010.08366}{{\ttfamily
  arXiv:2010.08366 [astro-ph.CO]}}.

\bibitem{Singal:2021crs}
A.~K. Singal, ``{Peculiar motion of Solar system from the Hubble diagram of
  supernovae Ia and its implications for cosmology},''
  \href{http://dx.doi.org/10.1093/mnras/stac1986}{{\em Mon. Not. Roy. Astron.
  Soc.} {\bfseries 515} no.~4, (2022) 5969--5980},
  \href{http://arxiv.org/abs/2106.11968}{{\ttfamily arXiv:2106.11968
  [astro-ph.CO]}}.

\bibitem{Tiwari:2015tba}
P.~Tiwari and A.~Nusser, ``{Revisiting the NVSS number count dipole},''
  \href{http://dx.doi.org/10.1088/1475-7516/2016/03/062}{{\em JCAP} {\bfseries
  03} (2016) 062}, \href{http://arxiv.org/abs/1509.02532}{{\ttfamily
  arXiv:1509.02532 [astro-ph.CO]}}.

\bibitem{Colin:2017juj}
J.~Colin, R.~Mohayaee, M.~Rameez, and S.~Sarkar, ``{High redshift radio
  galaxies and divergence from the CMB dipole},''
  \href{http://dx.doi.org/10.1093/mnras/stx1631}{{\em Mon. Not. Roy. Astron.
  Soc.} {\bfseries 471} no.~1, (2017) 1045--1055},
  \href{http://arxiv.org/abs/1703.09376}{{\ttfamily arXiv:1703.09376
  [astro-ph.CO]}}.

\bibitem{Rubart:2014lia}
M.~Rubart, D.~Bacon, and D.~J. Schwarz, ``{Impact of local structure on the
  cosmic radio dipole},''
  \href{http://dx.doi.org/10.1051/0004-6361/201423583}{{\em Astron. Astrophys.}
  {\bfseries 565} (2014) A111},
  \href{http://arxiv.org/abs/1402.0376}{{\ttfamily arXiv:1402.0376
  [astro-ph.CO]}}.

\bibitem{Dalang:2021ruy}
C.~Dalang and C.~Bonvin, ``{On the kinematic cosmic dipole tension},''
  \href{http://dx.doi.org/10.1093/mnras/stac726}{{\em Mon. Not. Roy. Astron.
  Soc.} {\bfseries 512} no.~3, (2022) 3895--3905},
  \href{http://arxiv.org/abs/2111.03616}{{\ttfamily arXiv:2111.03616
  [astro-ph.CO]}}.

\bibitem{vonHausegger:2024jan}
S.~von Hausegger, ``{The expected kinematic matter dipole is robust against
  source evolution},'' \href{http://dx.doi.org/10.1093/mnrasl/slae092}{{\em
  Mon. Not. Roy. Astron. Soc.} {\bfseries 535} no.~1, (2024) L49--L53},
  \href{http://arxiv.org/abs/2404.07929}{{\ttfamily arXiv:2404.07929
  [astro-ph.CO]}}.

\bibitem{vonHausegger:2024fcu}
S.~von Hausegger and C.~Dalang, ``{Redshift tomography of the kinematic matter
  dipole},'' \href{http://arxiv.org/abs/2412.13162}{{\ttfamily arXiv:2412.13162
  [astro-ph.CO]}}.

\bibitem{Ferreira:2020aqa}
P.~d.~S. Ferreira and M.~Quartin, ``{First Constraints on the Intrinsic CMB
  Dipole and Our Velocity with Doppler and Aberration},''
  \href{http://dx.doi.org/10.1103/PhysRevLett.127.101301}{{\em Phys. Rev.
  Lett.} {\bfseries 127} no.~10, (2021) 101301},
  \href{http://arxiv.org/abs/2011.08385}{{\ttfamily arXiv:2011.08385
  [astro-ph.CO]}}.

\bibitem{Ferreira:2021omv}
P.~d.~S. Ferreira and M.~Quartin, ``{Disentangling Doppler modulation,
  aberration and the temperature dipole in the CMB},''
  \href{http://dx.doi.org/10.1103/PhysRevD.104.063503}{{\em Phys. Rev. D}
  {\bfseries 104} no.~6, (2021) 063503},
  \href{http://arxiv.org/abs/2107.10846}{{\ttfamily arXiv:2107.10846
  [astro-ph.CO]}}.

\bibitem{daSilveiraFerreira:2024ddn}
P.~da~Silveira~Ferreira and V.~Marra, ``{Tomographic redshift dipole: testing
  the cosmological principle},''
  \href{http://dx.doi.org/10.1088/1475-7516/2024/09/077}{{\em JCAP} {\bfseries
  09} (2024) 077}, \href{http://arxiv.org/abs/2403.14580}{{\ttfamily
  arXiv:2403.14580 [astro-ph.CO]}}.

\bibitem{LIGOScientific:2017adf}
{\bfseries LIGO Scientific, Virgo, 1M2H, Dark Energy Camera GW-E, DES, DLT40,
  Las Cumbres Observatory, VINROUGE, MASTER} Collaboration, B.~P. Abbott {\em
  et~al.}, ``{A gravitational-wave standard siren measurement of the Hubble
  constant},'' \href{http://dx.doi.org/10.1038/nature24471}{{\em Nature}
  {\bfseries 551} no.~7678, (2017) 85--88},
  \href{http://arxiv.org/abs/1710.05835}{{\ttfamily arXiv:1710.05835
  [astro-ph.CO]}}.

\bibitem{Schutz1986}
B.~Schutz, ``Determining the hubble constant from gravitational wave
  observations,'' \href{http://dx.doi.org/10.1038/323310a0}{{\em Nature}
  {\bfseries 323} (1986) 310--311}. \url{https://doi.org/10.1038/323310a0}.

\bibitem{Finke:2021aom}
A.~Finke, S.~Foffa, F.~Iacovelli, M.~Maggiore, and M.~Mancarella, ``{Cosmology
  with LIGO/Virgo dark sirens: Hubble parameter and modified gravitational wave
  propagation},'' \href{http://dx.doi.org/10.1088/1475-7516/2021/08/026}{{\em
  JCAP} {\bfseries 08} (2021) 026},
  \href{http://arxiv.org/abs/2101.12660}{{\ttfamily arXiv:2101.12660
  [astro-ph.CO]}}.

\bibitem{Mastrogiovanni:2021wsd}
S.~Mastrogiovanni, K.~Leyde, C.~Karathanasis, E.~Chassande-Mottin, D.~A. Steer,
  J.~Gair, A.~Ghosh, R.~Gray, S.~Mukherjee, and S.~Rinaldi, ``{On the
  importance of source population models for gravitational-wave cosmology},''
  \href{http://dx.doi.org/10.1103/PhysRevD.104.062009}{{\em Phys. Rev. D}
  {\bfseries 104} no.~6, (2021) 062009},
  \href{http://arxiv.org/abs/2103.14663}{{\ttfamily arXiv:2103.14663 [gr-qc]}}.

\bibitem{Mastrogiovanni:2023emh}
S.~Mastrogiovanni, D.~Laghi, R.~Gray, G.~C. Santoro, A.~Ghosh, C.~Karathanasis,
  K.~Leyde, D.~A. Steer, S.~Perries, and G.~Pierra, ``{Joint population and
  cosmological properties inference with gravitational waves standard sirens
  and galaxy surveys},''
  \href{http://dx.doi.org/10.1103/PhysRevD.108.042002}{{\em Phys. Rev. D}
  {\bfseries 108} no.~4, (2023) 042002},
  \href{http://arxiv.org/abs/2305.10488}{{\ttfamily arXiv:2305.10488
  [astro-ph.CO]}}.

\bibitem{Mastrogiovanni:2023zbw}
S.~Mastrogiovanni, G.~Pierra, S.~Perri\`es, D.~Laghi, G.~Caneva~Santoro,
  A.~Ghosh, R.~Gray, C.~Karathanasis, and K.~Leyde, ``{ICAROGW: A python
  package for inference of astrophysical population properties of noisy,
  heterogeneous, and incomplete observations},''
  \href{http://dx.doi.org/10.1051/0004-6361/202347007}{{\em Astron. Astrophys.}
  {\bfseries 682} (2024) A167},
  \href{http://arxiv.org/abs/2305.17973}{{\ttfamily arXiv:2305.17973
  [astro-ph.CO]}}.

\bibitem{Gray:2019ksv}
R.~Gray {\em et~al.}, ``{Cosmological inference using gravitational wave
  standard sirens: A mock data analysis},''
  \href{http://dx.doi.org/10.1103/PhysRevD.101.122001}{{\em Phys. Rev. D}
  {\bfseries 101} no.~12, (2020) 122001},
  \href{http://arxiv.org/abs/1908.06050}{{\ttfamily arXiv:1908.06050 [gr-qc]}}.

\bibitem{Gray:2021sew}
R.~Gray, C.~Messenger, and J.~Veitch, ``{A pixelated approach to galaxy
  catalogue incompleteness: improving the dark siren measurement of the Hubble
  constant},'' \href{http://dx.doi.org/10.1093/mnras/stac366}{{\em Mon. Not.
  Roy. Astron. Soc.} {\bfseries 512} no.~1, (2022) 1127--1140},
  \href{http://arxiv.org/abs/2111.04629}{{\ttfamily arXiv:2111.04629
  [astro-ph.CO]}}.

\bibitem{Gray:2023wgj}
R.~Gray {\em et~al.}, ``{Joint cosmological and gravitational-wave population
  inference using dark sirens and galaxy catalogues},''
  \href{http://dx.doi.org/10.1088/1475-7516/2023/12/023}{{\em JCAP} {\bfseries
  12} (2023) 023}, \href{http://arxiv.org/abs/2308.02281}{{\ttfamily
  arXiv:2308.02281 [astro-ph.CO]}}.

\bibitem{LIGOScientific:2021aug}
{\bfseries LIGO Scientific, Virgo, KAGRA} Collaboration, R.~Abbott {\em
  et~al.}, ``{Constraints on the Cosmic Expansion History from
  GWTC\textendash{}3},'' \href{http://dx.doi.org/10.3847/1538-4357/ac74bb}{{\em
  Astrophys. J.} {\bfseries 949} no.~2, (2023) 76},
  \href{http://arxiv.org/abs/2111.03604}{{\ttfamily arXiv:2111.03604
  [astro-ph.CO]}}.

\bibitem{Stiskalek:2020wbj}
R.~Stiskalek, J.~Veitch, and C.~Messenger, ``{Are stellar--mass binary black
  hole mergers isotropically distributed?},''
  \href{http://dx.doi.org/10.1093/mnras/staa3613}{{\em Mon. Not. Roy. Astron.
  Soc.} {\bfseries 501} no.~1, (2021) 970--977},
  \href{http://arxiv.org/abs/2003.02919}{{\ttfamily arXiv:2003.02919
  [astro-ph.HE]}}.

\bibitem{Essick:2022slj}
R.~Essick, W.~M. Farr, M.~Fishbach, D.~E. Holz, and E.~Katsavounidis,
  ``{(An)isotropy measurement with gravitational wave observations},''
  \href{http://dx.doi.org/10.1103/PhysRevD.107.043016}{{\em Phys. Rev. D}
  {\bfseries 107} no.~4, (2023) 043016},
  \href{http://arxiv.org/abs/2207.05792}{{\ttfamily arXiv:2207.05792
  [astro-ph.HE]}}.

\bibitem{Kashyap:2022ibx}
G.~Kashyap, N.~K. Singh, K.~S. Phukon, S.~Caudill, and P.~Jain, ``{Dipole
  anisotropy in gravitational wave source distribution},''
  \href{http://dx.doi.org/10.1088/1475-7516/2023/06/042}{{\em JCAP} {\bfseries
  06} (2023) 042}, \href{http://arxiv.org/abs/2204.07472}{{\ttfamily
  arXiv:2204.07472 [astro-ph.CO]}}.

\bibitem{Punturo:2010zz}
M.~Punturo {\em et~al.}, ``{The Einstein Telescope: A third-generation
  gravitational wave observatory},''
  \href{http://dx.doi.org/10.1088/0264-9381/27/19/194002}{{\em Class. Quant.
  Grav.} {\bfseries 27} (2010) 194002}.

\bibitem{Branchesi:2023mws}
M.~Branchesi {\em et~al.}, ``{Science with the Einstein Telescope: a comparison
  of different designs},''
  \href{http://dx.doi.org/10.1088/1475-7516/2023/07/068}{{\em J. Cosmol.
  Astropart. Phys.} {\bfseries 07} (2023) 068},
  \href{http://arxiv.org/abs/2303.15923}{{\ttfamily arXiv:2303.15923 [gr-qc]}}.

\bibitem{Evans:2021gyd}
M.~Evans {\em et~al.}, ``{A Horizon Study for Cosmic Explorer: Science,
  Observatories, and Community},''
  \href{http://arxiv.org/abs/2109.09882}{{\ttfamily arXiv:2109.09882
  [astro-ph.IM]}}.

\bibitem{Srivastava:2022slt}
V.~Srivastava, D.~Davis, K.~Kuns, P.~Landry, S.~Ballmer, M.~Evans, E.~D. Hall,
  J.~Read, and B.~S. Sathyaprakash, ``{Science-driven Tunable Design of Cosmic
  Explorer Detectors},'' \href{http://dx.doi.org/10.3847/1538-4357/ac5f04}{{\em
  Astrophys. J.} {\bfseries 931} no.~1, (2022) 22},
  \href{http://arxiv.org/abs/2201.10668}{{\ttfamily arXiv:2201.10668 [gr-qc]}}.

\bibitem{Evans:2023euw}
M.~Evans {\em et~al.}, ``{Cosmic Explorer: A Submission to the NSF MPSAC ngGW
  Subcommittee},'' \href{http://arxiv.org/abs/2306.13745}{{\ttfamily
  arXiv:2306.13745 [astro-ph.IM]}}.

\bibitem{amaroseoane2017}
P.~Amaro-Seoane {\em et~al.}, ``Laser interferometer space antenna,'' 2017.
\newblock \url{https://arxiv.org/abs/1702.00786}.

\bibitem{Bayle:2022hvs}
J.-B. Bayle, B.~Bonga, C.~Caprini, D.~Doneva, M.~Muratore, A.~Petiteau,
  E.~Rossi, and L.~Shao, ``{Overview and progress on the Laser Interferometer
  Space Antenna mission},''
  \href{http://dx.doi.org/10.1038/s41550-022-01847-0}{{\em Nature Astron.}
  {\bfseries 6} no.~12, (2022) 1334--1338}.

\bibitem{Mastrogiovanni:2022nya}
S.~Mastrogiovanni, C.~Bonvin, G.~Cusin, and S.~Foffa, ``{Detection and
  estimation of the cosmic dipole with the einstein telescope and cosmic
  explorer},'' \href{http://dx.doi.org/10.1093/mnras/stad430}{{\em Mon. Not.
  Roy. Astron. Soc.} {\bfseries 521} no.~1, (2023) 984--994},
  \href{http://arxiv.org/abs/2209.11658}{{\ttfamily arXiv:2209.11658
  [astro-ph.CO]}}.

\bibitem{Grimm:2023tfl}
N.~Grimm, M.~Pijnenburg, S.~Mastrogiovanni, C.~Bonvin, S.~Foffa, and G.~Cusin,
  ``{Combining chirp mass, luminosity distance, and sky localization from
  gravitational wave events to detect the cosmic dipole},''
  \href{http://dx.doi.org/10.1093/mnras/stad3034}{{\em Mon. Not. Roy. Astron.
  Soc.} {\bfseries 526} no.~3, (2023) 4673--4689},
  \href{http://arxiv.org/abs/2309.00336}{{\ttfamily arXiv:2309.00336
  [astro-ph.CO]}}.

\bibitem{Cousins:2024bhk}
B.~Cousins, A.~Dhani, B.~S. Sathyaprakash, and N.~Yunes, ``{Finding cosmic
  anisotropy with networks of next-generation gravitational-wave detectors},''
  \href{http://arxiv.org/abs/2406.15550}{{\ttfamily arXiv:2406.15550 [gr-qc]}}.

\bibitem{Cai:2017aea}
R.-G. Cai, T.-B. Liu, X.-W. Liu, S.-J. Wang, and T.~Yang, ``{Probing cosmic
  anisotropy with gravitational waves as standard sirens},''
  \href{http://dx.doi.org/10.1103/PhysRevD.97.103005}{{\em Phys. Rev. D}
  {\bfseries 97} no.~10, (2018) 103005},
  \href{http://arxiv.org/abs/1712.00952}{{\ttfamily arXiv:1712.00952
  [astro-ph.CO]}}.

\bibitem{Cai:2019cfw}
R.-G. Cai, T.-B. Liu, S.-J. Wang, and W.-T. Xu, ``{Probing cosmic anisotropy
  with GW/FRB as upgraded standard sirens},''
  \href{http://dx.doi.org/10.1088/1475-7516/2019/09/016}{{\em JCAP} {\bfseries
  09} (2019) 016}, \href{http://arxiv.org/abs/1905.01803}{{\ttfamily
  arXiv:1905.01803 [astro-ph.CO]}}.

\bibitem{Cusin:2022cbb}
G.~Cusin and G.~Tasinato, ``{Doppler boosting the stochastic gravitational wave
  background},'' \href{http://dx.doi.org/10.1088/1475-7516/2022/08/036}{{\em
  JCAP} {\bfseries 08} no.~08, (2022) 036},
  \href{http://arxiv.org/abs/2201.10464}{{\ttfamily arXiv:2201.10464
  [astro-ph.CO]}}.

\bibitem{Tasinato:2023zcg}
G.~Tasinato, ``{Kinematic anisotropies and pulsar timing arrays},''
  \href{http://dx.doi.org/10.1103/PhysRevD.108.103521}{{\em Phys. Rev. D}
  {\bfseries 108} no.~10, (2023) 103521},
  \href{http://arxiv.org/abs/2309.00403}{{\ttfamily arXiv:2309.00403 [gr-qc]}}.

\bibitem{Cruz:2024svc}
N.~M.~J. Cruz, A.~Malhotra, G.~Tasinato, and I.~Zavala, ``{Measuring kinematic
  anisotropies with pulsar timing arrays},''
  \href{http://dx.doi.org/10.1103/PhysRevD.110.063526}{{\em Phys. Rev. D}
  {\bfseries 110} no.~6, (2024) 063526},
  \href{http://arxiv.org/abs/2402.17312}{{\ttfamily arXiv:2402.17312 [gr-qc]}}.

\bibitem{Cruz:2024diu}
N.~M.~J. Cruz, A.~Malhotra, G.~Tasinato, and I.~Zavala, ``{Astrometry meets
  Pulsar Timing Arrays: Synergies for Gravitational Wave Detection},''
  \href{http://arxiv.org/abs/2412.14010}{{\ttfamily arXiv:2412.14010
  [astro-ph.CO]}}.

\bibitem{Singer:2016eax}
L.~P. Singer {\em et~al.}, ``{Going the Distance: Mapping Host Galaxies of LIGO
  and Virgo Sources in Three Dimensions Using Local Cosmography and Targeted
  Follow-up},'' \href{http://dx.doi.org/10.3847/2041-8205/829/1/L15}{{\em
  Astrophys. J. Lett.} {\bfseries 829} no.~1, (2016) L15},
  \href{http://arxiv.org/abs/1603.07333}{{\ttfamily arXiv:1603.07333
  [astro-ph.HE]}}.

\bibitem{Nishizawa:2016ood}
A.~Nishizawa, ``{Measurement of Hubble constant with stellar-mass binary black
  holes},'' \href{http://dx.doi.org/10.1103/PhysRevD.96.101303}{{\em Phys. Rev.
  D} {\bfseries 96} no.~10, (2017) 101303},
  \href{http://arxiv.org/abs/1612.06060}{{\ttfamily arXiv:1612.06060
  [astro-ph.CO]}}.

\bibitem{Yu:2020vyy}
J.~Yu, Y.~Wang, W.~Zhao, and Y.~Lu, ``{Hunting for the host galaxy groups of
  binary black holes and the application in constraining Hubble constant},''
  \href{http://dx.doi.org/10.1093/mnras/staa2465}{{\em Mon. Not. Roy. Astron.
  Soc.} {\bfseries 498} no.~2, (2020) 1786--1800},
  \href{http://arxiv.org/abs/2003.06586}{{\ttfamily arXiv:2003.06586
  [astro-ph.CO]}}.

\bibitem{Borhanian:2020vyr}
S.~Borhanian, A.~Dhani, A.~Gupta, K.~G. Arun, and B.~S. Sathyaprakash, ``{Dark
  Sirens to Resolve the Hubble\textendash{}Lema\^\i{}tre Tension},''
  \href{http://dx.doi.org/10.3847/2041-8213/abcaf5}{{\em Astrophys. J. Lett.}
  {\bfseries 905} no.~2, (2020) L28},
  \href{http://arxiv.org/abs/2007.02883}{{\ttfamily arXiv:2007.02883
  [astro-ph.CO]}}.

\bibitem{Gupta:2022fwd}
I.~Gupta, ``{Using grey sirens to resolve the Hubble\textendash{}Lema\^\i{}tre
  tension},'' \href{http://dx.doi.org/10.1093/mnras/stad2115}{{\em Mon. Not.
  Roy. Astron. Soc.} {\bfseries 524} no.~3, (2023) 3537--3558},
  \href{http://arxiv.org/abs/2212.00163}{{\ttfamily arXiv:2212.00163 [gr-qc]}}.

\bibitem{Chen:2024gdn}
H.-Y. Chen, J.~M. Ezquiaga, and I.~Gupta, ``{Cosmography with next-generation
  gravitational wave detectors},''
  \href{http://dx.doi.org/10.1088/1361-6382/ad424f}{{\em Class. Quant. Grav.}
  {\bfseries 41} no.~12, (2024) 125004},
  \href{http://arxiv.org/abs/2402.03120}{{\ttfamily arXiv:2402.03120 [gr-qc]}}.

\bibitem{A_sharp_report}
P.~Fritschel {\em et~al.}, ``{Report of the LSC Post-O5 Study Group},'' 2024.
\newblock \url{https://dcc.ligo.org/LIGO-T2200287/public}.

\bibitem{Bonvin:2005ps}
C.~Bonvin, R.~Durrer, and M.~A. Gasparini, ``{Fluctuations of the luminosity
  distance},'' \href{http://dx.doi.org/10.1103/PhysRevD.85.029901}{{\em Phys.
  Rev. D} {\bfseries 73} (2006) 023523},
  \href{http://arxiv.org/abs/astro-ph/0511183}{{\ttfamily
  arXiv:astro-ph/0511183}}. [Erratum: Phys.Rev.D 85, 029901 (2012)].

\bibitem{Fishbach:2018edt}
M.~Fishbach, D.~E. Holz, and W.~M. Farr, ``{Does the Black Hole Merger Rate
  Evolve with Redshift?},''
  \href{http://dx.doi.org/10.3847/2041-8213/aad800}{{\em Astrophys. J. Lett.}
  {\bfseries 863} no.~2, (2018) L41},
  \href{http://arxiv.org/abs/1805.10270}{{\ttfamily arXiv:1805.10270
  [astro-ph.HE]}}.

\bibitem{Mandel:2018mve}
I.~Mandel, W.~M. Farr, and J.~R. Gair, ``{Extracting distribution parameters
  from multiple uncertain observations with selection biases},''
  \href{http://dx.doi.org/10.1093/mnras/stz896}{{\em Mon. Not. Roy. Astron.
  Soc.} {\bfseries 486} no.~1, (2019) 1086--1093},
  \href{http://arxiv.org/abs/1809.02063}{{\ttfamily arXiv:1809.02063
  [physics.data-an]}}.

\bibitem{Vitale:2020aaz}
S.~Vitale, D.~Gerosa, W.~M. Farr, and S.~R. Taylor, ``{Inferring the properties
  of a population of compact binaries in presence of selection effects},'' {\em
  Handbook of Gravitational Wave Astronomy} (7, 2020) p.45,
  \href{http://arxiv.org/abs/2007.05579}{{\ttfamily arXiv:2007.05579
  [astro-ph.IM]}}.

\bibitem{Borghi:2023opd}
N.~Borghi, M.~Mancarella, M.~Moresco, M.~Tagliazucchi, F.~Iacovelli,
  A.~Cimatti, and M.~Maggiore, ``{Cosmology and Astrophysics with Standard
  Sirens and Galaxy Catalogs in View of Future Gravitational Wave
  Observations},'' \href{http://dx.doi.org/10.3847/1538-4357/ad20eb}{{\em
  Astrophys. J.} {\bfseries 964} no.~2, (2024) 191},
  \href{http://arxiv.org/abs/2312.05302}{{\ttfamily arXiv:2312.05302
  [astro-ph.CO]}}.

\bibitem{Tagliazucchi:2025ofb}
M.~Tagliazucchi, M.~Moresco, N.~Borghi, and M.~Fiebig, ``{Accelerating the
  Standard Siren Method: Improved Constraints on Modified Gravitational Wave
  Propagation with Future Data},''
  \href{http://arxiv.org/abs/2504.02034}{{\ttfamily arXiv:2504.02034
  [astro-ph.CO]}}.

\bibitem{Dalya:2021ewn}
G.~D\'alya {\em et~al.}, ``{GLADE+~: an extended galaxy catalogue for
  multimessenger searches with advanced gravitational-wave detectors},''
  \href{http://dx.doi.org/10.1093/mnras/stac1443}{{\em Mon. Not. Roy. Astron.
  Soc.} {\bfseries 514} no.~1, (2022) 1403--1411},
  \href{http://arxiv.org/abs/2110.06184}{{\ttfamily arXiv:2110.06184
  [astro-ph.CO]}}.

\bibitem{Chen:2023wpj}
A.~Chen, R.~Gray, and T.~Baker, ``{Testing the nature of gravitational wave
  propagation using dark sirens and galaxy catalogues},''
  \href{http://dx.doi.org/10.1088/1475-7516/2024/02/035}{{\em JCAP} {\bfseries
  02} (2024) 035}, \href{http://arxiv.org/abs/2309.03833}{{\ttfamily
  arXiv:2309.03833 [gr-qc]}}.

\bibitem{Madau:2014bja}
P.~Madau and M.~Dickinson, ``{Cosmic Star Formation History},''
  \href{http://dx.doi.org/10.1146/annurev-astro-081811-125615}{{\em Ann. Rev.
  Astron. Astrophys.} {\bfseries 52} (2014) 415--486},
  \href{http://arxiv.org/abs/1403.0007}{{\ttfamily arXiv:1403.0007
  [astro-ph.CO]}}.

\bibitem{Callister:2020arv}
T.~Callister, M.~Fishbach, D.~Holz, and W.~Farr, ``{Shouts and Murmurs:
  Combining Individual Gravitational-Wave Sources with the Stochastic
  Background to Measure the History of Binary Black Hole Mergers},''
  \href{http://dx.doi.org/10.3847/2041-8213/ab9743}{{\em Astrophys. J. Lett.}
  {\bfseries 896} no.~2, (2020) L32},
  \href{http://arxiv.org/abs/2003.12152}{{\ttfamily arXiv:2003.12152
  [astro-ph.HE]}}.

\bibitem{KAGRA:2021duu}
{\bfseries KAGRA, VIRGO, LIGO Scientific} Collaboration, R.~Abbott {\em
  et~al.}, ``{Population of Merging Compact Binaries Inferred Using
  Gravitational Waves through GWTC-3},''
  \href{http://dx.doi.org/10.1103/PhysRevX.13.011048}{{\em Phys. Rev. X}
  {\bfseries 13} no.~1, (2023) 011048},
  \href{http://arxiv.org/abs/2111.03634}{{\ttfamily arXiv:2111.03634
  [astro-ph.HE]}}.

\bibitem{Karathanasis:2022hrb}
C.~Karathanasis, B.~Revenu, S.~Mukherjee, and F.~Stachurski, ``{GWSim: Python
  package for creating mock GW samples for different astrophysical populations
  and cosmological models of binary black holes},''
  \href{http://dx.doi.org/10.1051/0004-6361/202245216}{{\em Astron. Astrophys.}
  {\bfseries 677} (2023) A124},
  \href{http://arxiv.org/abs/2210.05724}{{\ttfamily arXiv:2210.05724
  [astro-ph.CO]}}. [Erratum: Astron.Astrophys. 682, C1 (2024)].

\bibitem{deSouza:2023ozp}
J.~M.~S. de~Souza and R.~Sturani, ``{GWDALI: A Fisher-matrix based software for
  gravitational wave parameter-estimation beyond Gaussian approximation},''
  \href{http://dx.doi.org/10.1016/j.ascom.2023.100759}{{\em Astron. Comput.}
  {\bfseries 45} (2023) 100759},
  \href{http://arxiv.org/abs/2307.10154}{{\ttfamily arXiv:2307.10154 [gr-qc]}}.

\bibitem{Pratten:2020ceb}
G.~Pratten {\em et~al.}, ``{Computationally efficient models for the dominant
  and subdominant harmonic modes of precessing binary black holes},''
  \href{http://dx.doi.org/10.1103/PhysRevD.103.104056}{{\em Phys. Rev. D}
  {\bfseries 103} no.~10, (2021) 104056},
  \href{http://arxiv.org/abs/2004.06503}{{\ttfamily arXiv:2004.06503 [gr-qc]}}.

\bibitem{1976ApJ...203..297S}
P.~{Schechter}, ``{An analytic expression for the luminosity function for
  galaxies.},'' \href{http://dx.doi.org/10.1086/154079}{{\em Astrophys. J.}
  {\bfseries 203} (Jan., 1976) 297--306}.

\bibitem{Longair:2008gba}
M.~S. Longair, \href{http://dx.doi.org/10.1007/978-3-540-73478-9}{{\em {Galaxy
  Formation}}}.
\newblock Astronomy and Astrophysics Library. Springer, Heidelberg, Germany,
  2008.

\bibitem{Poisson:1995ef}
E.~Poisson and C.~M. Will, ``{Gravitational waves from inspiraling compact
  binaries: Parameter estimation using second postNewtonian wave forms},''
  \href{http://dx.doi.org/10.1103/PhysRevD.52.848}{{\em Phys. Rev. D}
  {\bfseries 52} (1995) 848--855},
  \href{http://arxiv.org/abs/gr-qc/9502040}{{\ttfamily arXiv:gr-qc/9502040}}.

\bibitem{emcee}
D.~{Foreman-Mackey}, D.~W. {Hogg}, D.~{Lang}, and J.~{Goodman}, ``{emcee: The
  MCMC Hammer},'' \href{http://dx.doi.org/10.1086/670067}{{\em PASP} {\bfseries
  125} no.~925, (Mar., 2013) 306},
  \href{http://arxiv.org/abs/1202.3665}{{\ttfamily arXiv:1202.3665
  [astro-ph.IM]}}.

\bibitem{Nishizawa:2010xx}
A.~Nishizawa, A.~Taruya, and S.~Saito, ``{Tracing the redshift evolution of
  Hubble parameter with gravitational-wave standard sirens},''
  \href{http://dx.doi.org/10.1103/PhysRevD.83.084045}{{\em Phys. Rev. D}
  {\bfseries 83} (2011) 084045},
  \href{http://arxiv.org/abs/1011.5000}{{\ttfamily arXiv:1011.5000
  [astro-ph.CO]}}.

\bibitem{Nimonkar:2023pyt}
H.~Nimonkar and S.~Mukherjee, ``{Dependence of peculiar velocity on the host
  properties of the gravitational wave sources and its impact on the
  measurement of Hubble constant},''
  \href{http://dx.doi.org/10.1093/mnras/stad3256}{{\em Mon. Not. Roy. Astron.
  Soc.} {\bfseries 527} no.~2, (2023) 2152--2164},
  \href{http://arxiv.org/abs/2307.05688}{{\ttfamily arXiv:2307.05688
  [astro-ph.CO]}}.

\bibitem{Cusin:2024git}
G.~Cusin, C.~Pitrou, C.~Bonvin, A.~Barrau, and K.~Martineau, ``{Boosting
  gravitational waves: a review of kinematic effects on amplitude,
  polarization, frequency and energy density},''
  \href{http://dx.doi.org/10.1088/1361-6382/ad7ad0}{{\em Class. Quant. Grav.}
  {\bfseries 41} no.~22, (2024) 225006},
  \href{http://arxiv.org/abs/2405.01297}{{\ttfamily arXiv:2405.01297 [gr-qc]}}.

\end{thebibliography}\endgroup

\end{document}